\documentclass[11pt]{article}
\usepackage[utf8]{inputenc}
\usepackage{amsmath}
\usepackage{lipsum} 
\usepackage{fullpage}
\usepackage{mathrsfs}
\usepackage{svg}
\usepackage{xr}
\usepackage{amsthm,cite,url,graphicx,booktabs,lipsum,color,bm,caption,subcaption,soul}
\usepackage{pifont,tikz,paralist,multirow,amssymb}
\usepackage{xifthen}
\usepackage{enumerate}
\usepackage{titlesec}
\usepackage{arydshln}
\usepackage{natbib} 
\usepackage{color}
\usepackage{tikz}
\usepackage{multirow}

\newtheorem{theorem}{Theorem}

\newtheorem{proposition}{Proposition}
\newtheorem{assumption}{Assumption}

\newtheorem{condition}{Condition}
\usepackage{hyperref}
\hypersetup{
    colorlinks=true,
    linkcolor=blue,
    urlcolor=blue,
    citecolor=blue
}
\usetikzlibrary{positioning}

\allowdisplaybreaks
\newcommand{\V}[1]{\boldsymbol{#1}} 

\newtheorem{innercustomthm}{Assumption}
\newenvironment{customassump}[1]
{\renewcommand\theinnercustomthm{#1}\innercustomthm}
{\endinnercustomthm}
\makeatother

\def \P{\mathbb{P}} 
\def \V {\mathbb{V}}

\def \E {\mathbb{E}}

\makeatletter
\newcommand*{\indep}{%
 \mathbin{%
  \mathpalette{\@indep}{}%
 }%
}
\newcommand*{\nindep}{%
 \mathbin{
  \mathpalette{\@indep}{\not}
 }%
}
\newcommand*{\@indep}[2]{%
 \sbox0{$#1\perp\m@th$}
 \sbox2{$#1=$}
 \sbox4{$#1\vcenter{}$}
 \rlap{\copy0}
 \dimen@=\dimexpr\ht2-\ht4-.2pt\relax
 \kern\dimen@
 {#2}%
 \kern\dimen@
 \copy0 
}
\usepackage{xr}

\makeatletter
\newcommand*{\addFileDependency}[1]{
  \typeout{(#1)}
  \@addtofilelist{#1}
  \IfFileExists{#1}{}{\typeout{No file #1.}}
}
\makeatother

\newcommand*{\myexternaldocument}[1]{
    \externaldocument{#1}
    \addFileDependency{#1.tex}
    \addFileDependency{#1.aux}
}
 \usepackage{tocloft}
 \usepackage{setspace}
 \myexternaldocument{main}

\begin{document}

\hypersetup{linkcolor=black}
\title{\bf On the Comparative Analysis of Average Treatment Effects Estimation via Data Combination}
\author{Peng Wu\thanks{School of Mathematics and Statistics, Beijing Technology and Business University, \href{mailto:pengwu@btbu.edu.cn}{pengwu@btbu.edu.cn}}, ~Shanshan Luo\thanks{School of Mathematics and Statistics, Beijing Technology and Business University, \href{mailto:shanshanluo@btbu.edu.cn}{shanshanluo@btbu.edu.cn}}, ~and Zhi Geng\thanks{School of Mathematics and Statistics, Beijing Technology and Business University, \href{mailto:zhigeng@pku.edu.cn}{zhigeng@pku.edu.cn}} }

\date{} 
\maketitle 
\bigskip
\hypersetup{linkcolor=blue}


\bigskip
\bigskip
\bigskip
\begin{abstract} 
\noindent%
There is growing interest in exploring causal effects in target populations via data combination. However, most approaches are tailored to specific settings and lack comprehensive comparative analyses.  In this article, we focus on a typical scenario involving a source dataset and a target dataset. We first design six settings under covariate shift and conduct a comparative analysis by deriving the semiparametric efficiency bounds for the ATE in the target population. We then extend this analysis to six new settings that incorporate both covariate shift and posterior drift. Our study uncovers the key factors that influence efficiency gains and the ``effective sample size" when combining two datasets, with a particular emphasis on the roles of the variance ratio of potential outcomes between datasets and the derivatives of the posterior drift function. To the best of our knowledge, this is the first paper that explicitly explores the role of the posterior drift functions in causal inference. 
Additionally,  we also propose novel methods for conducting sensitivity analysis to address violations of transportability between the two datasets. We empirically validate our findings by constructing locally efficient estimators and conducting extensive simulations. We demonstrate the proposed methods in two real-world applications.  

 \end{abstract} 
 
\vfill

\bigskip
 
\newpage

\section{Introduction} \label{sec:intro}
The availability of multiple datasets collected from diverse populations under different designs has opened up new opportunities for causal inference~\citep{stuart2011use, Bareinboim-Pearl2016, lesko2017generalizing, Colnet-etal2023}.   However, it also introduces challenges when making inferences about a specific target population, primarily due to the potential incomparability between different datasets~\citep{Wu2023Transfer,Yang-etal2023}. Specifically, consider a typical scenario of two datasets: a source dataset and a target dataset.  One common challenge is the absence of critical variables in the target dataset, such as long-term outcome data or treatment assignment information~\citep{athey-etal2019, He-etal2020, hu2023longterm}. Moreover, there is often heterogeneity in the data distribution between the source and target datasets. For example, the source dataset may come from randomized controlled trials (RCT), while the target dataset comes from observational studies. Additionally, unmeasured confounding can be a concern in the target dataset if it is collected from observational studies~\citep{Yang-Ding2020}.

Numerous studies have demonstrated the necessity of combining datasets from different designs to perform causal inference~\citep{li2023improving, Colnet-etal2023, Humermund-Bareinboim2020}. Under reasonable assumptions, this strategy not only helps to address the challenges mentioned above but also significantly improves estimation efficiency, compared to methods using a single dataset. Due to the potential and power of data combination, there has been a surge of related work, dispersed across different disciplines and using slightly different terminology, including generalizability~\citep{stuart2011use, Buchanan-etal2018, Dahabreh-etal2019, Cinelli-Pearl2021}, transportability~\citep{HernanVanderWeele2011, Rudolph2017, Westreich-etal2017, Degtiar-Rose2023}, external validity~\citep{Westreich-etal2017},  data fusion~\citep{Bareinboim-Pearl2016, Humermund-Bareinboim2020}, and so on.

Despite the popularity and theoretical appeal of data combination methods, most of the existing approaches are tailored to specific settings, and there is a notable lack of comprehensive comparative analyses across different settings. In data combination methods, it is well-known that  collecting additional auxiliary datasets and imposing specific assumptions are indispensable to ensure identifiability or improve estimation efficiency, in contrast to methods relying on a single dataset~\citep{Colnet-etal2023, Humermund-Bareinboim2020}. 
Several questions naturally arise: what are the respective roles of these auxiliary datasets and specific assumptions, and which holds greater importance – the datasets or the assumptions? 
  Do covariates, treatments, and outcomes in the auxiliary datasets contribute meaningfully, or does their utility depend on the assumptions? Furthermore, under what assumptions do these variables come into play a substantial role? Addressing these questions necessitates comparative analyses under varying settings.



In this article, we aim to answer these questions by focusing on a typical scenario that involves a source dataset and a target dataset. \emph{The key lies in (a) designing various settings, each with different levels of information in terms of data and assumptions, and (b) conducting a comprehensive comparative analysis across these settings.}  We assume that the source dataset is independent of the target dataset, with the latter being representative of the target population. Also,  we allow for significant differences in the distributions of the two datasets, including both covariate shift and posterior drift  \citep{TonyCai2021}. 
Covariate shift refers to differences in the distribution of covariates between the two datasets, while posterior drift refers to differences in the conditional distribution of potential outcomes given the covariates between the two datasets. 


 Toward the first goal (a), we first examine four different cases about the target dataset,  each characterized by varying levels of information: (i) only covariates are available; (ii) both the covariates and treatment are available; (iii) both the covariates and outcome are available; (iv) all covariates, treatment, and outcome are available.  The source dataset is assumed to have complete information on covariates, treatment, and outcome. Then,  under the case of covariate shift, we consider three sets of identifiability assumptions for treatment effects in the target population, see Section \ref{sec:identifiablity} for more details. Finally, we design six distinct settings to explore. 
The six settings compared in this paper encompass various scenarios from the existing literature on causal inference through data combination, 
 For instance, our setting I aligns with the setting discussed in \cite{Hotz-etal2005, Cole-Stuart2010, stuart2011use, Tipton2013, Hartman-etal2015, Kern2016, lesko2017generalizing, Buchanan-etal2018, Li-etal2022, Li-etal2023}, and our setting IV corresponds to the setting of \cite{Rudolph2017, Kallus-Puli2018, Wu-etal2012-CCLR, Yang-etal2022-arXiv, Hatt-etal2022-arXiv}, albeit focusing on different target estimands.
Furthermore, we expand the six settings to encompass a broader scenario involving both covariate shift and posterior drift, see Sections \ref{sec:identifiablity} and \ref{sec:sen} for more details. These expanded settings are rarely explored in previous literature, extending the settings of \citet{Dahabreh-etal2019, Dahabreh-etal2020, Lee-etal2021} and incorporating findings from \cite{li2023improving}.

For the second goal (b), we perform a comprehensive efficiency comparative analysis across the six settings. Initially, for each setting, we calculate the efficient influence {function} and the associated semiparametric efficiency bound for the average treatment effect (ATE) in the target population. Subsequently, we compare these semiparametric efficiency bounds among the six settings, thereby theoretically quantifying the impact of various levels of information on the estimation of ATE in the target population. Under covariate shift, the comparative analysis reveal the key factors that contribute to efficiency gains/losses across these settings, including the proportion of the source dataset relative to the entire dataset, the variances of potential outcomes given covariates in the two datasets, the proportion of controls in the target data relative to all controls, the proportion of treated units in the target data relative to all treated units. Moreover, under both covariate shift and posterior drift, the key factors also include the first-order derivative of the posterior drift function and the variance ratio of potential outcomes between the two datasets. Interestingly, the product of the two additional factors quantifies the ``effect sample size ratio" in the target dataset, which can be interpreted as the optimal transfer rate. To the best of our knowledge, this is the first paper that explicitly reveals the crucial  role of the posterior drift function.

In summary, this article makes four major contributions. First, we design six settings under covariate shift and perform a comparative analysis to reveal key factors influencing estimation efficiency, see Sections \ref{sec:identifiablity} and \ref{sec:methodology} for details. 
Second, we extend the comparative analysis to scenarios where both covariate shift and posterior drift are present. This further reveals the critical role of the variance ratio of potential outcomes between  two datasets and the first-order derivative of the posterior drift function in estimation efficiency, as well as their impact on the ``effective sample size", see Section \ref{sec:sen} for details. 
Third, we propose double machine learning estimators for the ATE in the target population under both covariate shift and posterior drift, and show their large sample properties.
  Fourth, we propose a novel sensitivity analysis approach to assess the robustness of the proposed method regarding the transportability between the two datasets.
Additionally, we extend the comparative analysis to other estimands, including ATE in the source population, and ATE on the treated (ATT) in both source and target populations. 

The rest of this paper is organized as follows.  In Section \ref{ssec:setup}, we introduce the notation and basic setup. 
In Section \ref{sec:identifiablity}, we design twelve different settings by performing a comparative analysis of identifiability assumptions.  
In Section \ref{sec:methodology}, we consider the efficiency gains and losses among the designed six settings under the case of covariate shift.  In Section \ref{sec:sen}, we extend the efficiency analysis to another six settings, incorporating both covariate shift and posterior drift.   
  In Section \ref{sec-estimation}, we propose nonparametric machine learning estimators for the ATE in the target population under covariate shift alone and in cases where both covariate shift and posterior drift are present. Section \ref{sec:sen2} presents the proposed sensitivity analysis approach.  
Section \ref{sim-studies} presents extensive simulation studies to evaluate the finite sample performance of the proposed estimators. In Section \ref{sec:job-app}, we apply the proposed methodology to two real-world applications. In Sections \ref{sec:extension-ATE-source} and \ref{sec:ext-ATT}, we extend the comparative analysis to ATE in the source population, as well as ATE on the treated (ATT) in both source and target populations.  
We conclude with a brief discussion in Section  \ref{sec:conclusion}.


\section{Basic Setup} \label{ssec:setup}
Let $X$ be a vector of pre-treatment covariates, $Y$ be the outcome of interest, and $A$ be a binary treatment, where $A=1$ represents receiving treatment and $A = 0$ otherwise.  We adopt the potential outcome framework  to define causal effects~\citep{neyman1923, Rubin1974}. Under the stable unit treatment value  assumption~\citep{Rubin1980, rubin1990comment}, we denote the potential outcomes under treatment arms $0$ and $1$ as $Y(0)$ and $Y(1)$, respectively. By the consistency assumption, 
the observed outcome $Y= Y(A)= AY(1) + (1-A)Y(0)$. 

In this article, we consider a typical scenario of two datasets: a source dataset and a target dataset, and adopt the non-nested design where the two datasets are independent~\citep{dahabreh2021study}.  
Let $G$ be the indicator of the data source, where $G=1$ represents the source population and $G=0$ represents the target population.  
Under a superpopulation model $\P$, we assume that the data $\{(X_i,A_i,Y_i(0),Y_i(1),G_i),i=1,\ldots,n\}$ are independent and identically distributed for all units in the target and source datasets. 
Let $\E$ be the expectation operator of $\P$, then the ATE in the target population is denoted as $\tau = \E\{Y(1) - Y(0) \mid G = 0\}$.  
Let $q = \P(G = 1)$ be the probability of a unit belonging to the source population.  We allow for significant differences in the distributions of the two datasets, i.e., $\P(\cdot\mid G=1)\neq \P(\cdot\mid G=0)$.


The source dataset is assumed to have complete information on covariates, treatment, and outcome. Considering that researchers often cannot collect the complete information for the target population in practice, we examine four different scenarios of data structure about the target dataset as follows.   
\begin{enumerate}
      \item[(i)]  Only covariates data are available. This is often the case, for example, when researchers are trying to assess the efficacy of a drug that has not been previously tested in the target population. More real-world examples can be found in \cite{Dahabreh-etal2019, Dahabreh-etal2020, He-etal2020, Li-etal2022,  Li-etal2023}. 
      

     \item[(ii)] Covariates and treatment data are available, but outcome data are fully missing. This occurs, for example, when the outcome cannot be observed due to limited follow-up time, but a treatment has been assigned or a treatment rule has been designed in the target data, see \cite{imbens2022long, hu2023longterm} for practical instances.  %

     
     \item[(iii)] Covariates and outcome data are available, but treatment data are fully missing. This situation may occur when the outcome has been collected historically, but the treatment was not recorded. For more examples, please refer to \cite{athey-etal2019}.

     \item[(iv)] All covariates, treatment, and outcome data are available. For example, the target data comes from observational studies and is affected by unobserved confounders. For practical examples, please refer to  \cite{Wu-etal2012-CCLR, Yang-etal2022-arXiv, Yang-etal2023}. 
\end{enumerate}

Without additional assumptions, none of the four scenarios described above ensures the identifiability of $\tau$. In Section \ref{sec:identifiablity}, we explore the necessary identifiability assumptions for $\tau$.

\section{Assumptions Comparison for Identifiability}
\label{sec:identifiablity}
We initiate our exploration by considering two common assumptions, namely Assumptions \ref{assump1} and \ref{assump2}, which  are widely used in the literature on causal inference via data combination, see e.g., \cite{Dahabreh-etal2019, Dahabreh-etal2020, Li-etal2022,  Li-etal2023, Colnet-etal2023}.  

\begin{assumption}[Internal validity of the source data]  \label{assump1}   (i) $A \indep \{ Y(0), Y(1) \} \mid X, G = 1$, and (ii) $0 < e_1(X) := \P(A=1\mid X , G=1) < 1$ for all $X  $.   
\end{assumption}   
Assumption \ref{assump1} rules out unmeasured confounding between the treatment and outcome in the source data, and it holds in either randomized controlled trials or unconfounded
observational studies.  Assumption \ref{assump1} is sufficient to identify causal effects in the source population but not in the target population. Therefore,  we proceed to invoke Assumption \ref{assump2}.


\begin{assumption} \label{assump2} 
(i) Transportability: $G \indep \{ Y(0), Y(1)\} \mid X$  for all $X  $, and   (ii) Positivity of sampling score:  $0 < \pi(X)  := \P(G=1\mid X) < 1$  for all $X  $.  
\end{assumption}   
Assumption \ref{assump2} implies that the sampling mechanism of $G$ in the whole population is strongly ignorable~\citep{Li-etal2023}, which ensures the transportability of the conditional average treatment effect from the source population to the target population, i.e., 
\begin{align*}
  \tau(X):=  \E\{Y(1) - Y(0)\mid X , G=1\} = \E\{Y(1) - Y(0)\mid X  , G=0\}.
\end{align*}  
Under Assumption \ref{assump1}, $\tau(x)$ is identified as $\mu_1(x) - \mu_0(x)$, where $\mu_a(x) = \E(Y\mid X = x, A = a, G = 1 )$ for $a = 0, 1$. Then, if we have access to the covariates $X$ in the target data, Assumptions \ref{assump1}-\ref{assump2} imply that $\tau$ can be identified as $\tau = \E\{\mu_1(X) - \mu_0(X)  \mid G = 0\}.$
  Assumption \ref{assump2} also restricts that the conditional variance of potential outcomes remains equal between the two populations, denoted by $\text{Var}\{Y(a)\mid X, G=1\} = \text{Var}\{Y(a)\mid X, G=0\}$. This restriction will influence the efficiency of $\tau$ (see Theorem \ref{thm5} for details).

Let $\tilde\mu_a(X) = \E\{Y(a) \mid X, G=0\}$ for $a = 0,1$. In addition to Assumption \ref{assump2}, another commonly used but weaker assumption is the mean exchangeability~\citep{dahabreh2019relation, dahabreh2021study, li2023improving}, which states that $\tilde\mu_a(X) = \mu_a(X)$ for $a=0,1$. 
To further relax Assumption \ref{assump2} and explore the impact of posterior drift, we extend the mean exchangeability using two general posterior drift functions $\psi_{0}(\cdot)$ and $\psi_{1}(\cdot)$.

\begin{customassump}{2$^{\ast}$} \label{assump5} 
Assume ${\tilde \mu}_0(X)=\psi_{0} \{\mu_0(X)\}$ and ${\tilde \mu}_1(X)=\psi_{1} \{\mu_1(X)\}$, where $\psi_{0} (\cdot)$ and $\psi_{1} (\cdot)$ are two known differentiable functions.
\end{customassump}
Under Assumption \ref{assump1}, Assumption \ref{assump5} is notably weaker than Assumption \ref{assump2}, as it allows for variability in the conditional distributions of potential outcomes between the target population \(G=0\) and the source population \(G=1\), thus permitting posterior drift. In Assumption \ref{assump5}, the drift functions \(\psi_{0}(\cdot)\) and \(\psi_{1}(\cdot)\) determine the scale of deviation between the two conditional expectations. For example, under the   mean exchangeability assumption \citep{li2023improving}, \(\psi_{0}(u) = \psi_{1}(u) = u\). Additionally, if \(\psi_{a}(\cdot)\) is a monotonic function, it implies that \(\tilde \mu_a(X)\) and \(\mu_a(X)\) preserve their relative rank order. For any given \(\psi_{0}(\cdot)\) and \(\psi_{1}(\cdot)\), \(\tau\) can be identified as \( \tau = \E[\psi_{1}\{\mu_1(X)\} - \psi_{0}\{\mu_0(X)\} \mid G=0] \).


We emphasize that Assumptions \ref{assump1}-\ref{assump2} or Assumptions \ref{assump1}-\ref{assump5} do not require unconfoundedness (Assumption \ref{assump3}(i) below) to hold in the target population, allowing for the existence of certain unmeasured confounders~\citep{Kallus-Puli2018, Wu-etal2012-CCLR, Yang-etal2022-arXiv}. 
Consequently, $\mu_a(x)$ may not necessarily equal $\E(Y\mid X = x, A = a, G = 0)$. 
Also, Assumptions \ref{assump1}-\ref{assump2} do not restrict the propensity score in the target population to be equal to those in the source population; they do not even necessitate the positivity assumption of the propensity score to hold in the target population. 

To further explore the impact of the unconfoundedness and positivity assumptions in the target population, we present them below for comparison. 
\begin{assumption}[Internal validity of the target data]  \label{assump3} 
(i)  $A \indep \{Y(0), Y(1)\} \mid X, G=0$, and (ii) 
$0 < e_0(X) := \P(A=1\mid X , G=0) <1$ for all $X$. 
\end{assumption}

If the covariates, treatment, and outcome are available in the target population, Assumption \ref{assump3} is sufficient to 
identify $\tau$. 
In such a case, the source data is helpful to enhance the estimation efficiency of $\tau$ (see Proposition \ref{prop1} in Section \ref{sec:methodology}). It is noteworthy that under Assumptions \ref{assump1}-\ref{assump3}, the two propensity scores  $e_1(X)$ and $e_0(X)$ may not be equal. This means that the treatment assignment mechanisms between the source population and the target population are allowed to be different, which accommodates a variety of application scenarios, such as the source data being experimental (or observational) data and the target data being another experimental (or observational) data. 

In some scenarios, the target data may contain only control units and not any treated units~\citep{li2023improving}. 
To incorporate such scenarios, we introduce Assumption \ref{assump4} as an alternative to Assumption \ref{assump3}.

\begin{assumption}[No treated units in the target data] \label{assump4}    $A = 0$ if $G=0$ and thus $Y=Y(0)$ for each observation  in the target population.
\end{assumption}

 Clearly, Assumption \ref{assump4} implies $e_0(X) = 0$ for all $X$. Moreover,   
under Assumption \ref{assump4}, with $A$ being constant, the conditional independence  $ A \indep \{Y(0), Y(1)\} \mid X, G=0$  holds naturally. Thus, Assumption \ref{assump4} can be regarded as a special case of Assumption \ref{assump3}(i).

\begin{table}[t]\centering
\caption{Twelve settings for comparative analysis.}
\label{tab:6settings}\resizebox{0.9485\textwidth}{!}{
\begin{tabular}{clccccccccccc} 
\toprule\addlinespace[1mm]
Observed Data Structure    &  &  & \multicolumn{3}{c}{Without Posterior Drift} &  && & \multicolumn{3}{c}{With  Posterior Drift}  \\ 
\cline{4-6}\cline{10-12} 
  in Target Population &  &  & Setting &  & Assumptions            &     &    &  & Setting &  & Assumptions                   \\ 
\cline{1-1}\cline{4-6}\cline{10-12}
$X$                       &  &  & I       &  &   Assumptions \ref{assump1}, \ref{assump2}                  &  &  &   & I$^\ast$    &  & Assumptions \ref{assump1}, \ref{assump5}               \\
$(A,X)$                   &  &  & II      &  &  Assumptions \ref{assump1}, \ref{assump2}                 &  &   &  &   II$^\ast$  &  & Assumptions \ref{assump1}, \ref{assump5}                 \\
$(X,Y)$                   &  &  & III     &  &  Assumptions \ref{assump1}, \ref{assump2}                 &  & &    &   III$^\ast$  &  & Assumptions \ref{assump1}, \ref{assump5}              \\
$(A,X,Y)$                 &  &  & IV      &  &  Assumptions \ref{assump1}, \ref{assump2}                   &  & & &    IV$^\ast$    &  & Assumptions \ref{assump1}, \ref{assump5}                \\
$(A=0, X,Y)$              &  &  & V       &  &  Assumptions \ref{assump1}, \ref{assump2}, \ref{assump4}               &  &  &   &   V$^\ast$   &  &Assumptions \ref{assump1}, \ref{assump5}, \ref{assump4}               \\
$(A,X,Y)$                 &  &  & VI      &  &  Assumptions \ref{assump1}, \ref{assump2}, \ref{assump3}           &  &   & &  VI$^\ast$     &  & Assumptions \ref{assump1}, \ref{assump5}, \ref{assump3}                \\
\bottomrule
\end{tabular}}
\label{tab:scenarios}
\end{table} 

Combining the four target data structures in Section \ref{ssec:setup} with identifiability Assumptions \ref{assump1}-\ref{assump4}, we have designed twelve settings for comparative analysis. Each setting encompasses different levels of information across both datasets and assumptions. Table \ref{tab:scenarios} summarizes these settings, where the second column lists the scenarios under Assumption \ref{assump2}, and the fourth column lists the scenarios under Assumption \ref{assump5}. Both Assumption \ref{assump2} and Assumption \ref{assump5} allow for \(\P(X\mid G=1) \neq \P(X\mid G=0)\), indicating the presence of covariate shift, while the former requires \(\P(Y(a)\mid X, G=1) = \P(Y(a)\mid X, G=0)\), indicating identical conditional distributions between the two datasets, which the latter does not require. To simplify the exposition, we will explore the relative efficiency of these settings in Table \ref{tab:scenarios} in Sections \ref{sec:methodology} and \ref{sec:sen}, respectively.

\section{Eﬀiciency Comparison under Covariate Shift}
\label{sec:methodology}


In this section, we  perform a comprehensive efficiency comparative analysis for the first six settings outlined in the second column of  Table \ref{tab:scenarios}, with a focus on the ATE in the target population. 
We employ the semiparametric theory to conduct the efficiency analysis, as it provides a principled way to quantify how various levels of information in observed data affect the estimation of causal parameters of interest~\citep{vdv-1998, Tsiatis-2006}.  

 
Theorem \ref{thm1} below gives the efficient influence functions (EIFs)  and the corresponding semiparametric efficiency bounds for $\tau$ under settings from I to VI, serving as the theoretical foundation for subsequent comparisons of estimation efficiency.

\begin{theorem}[EIFs of $\tau$] \label{thm1} 
The following statements hold: 
\begin{itemize}
    \item[(a)] the efficient influence function of $\tau$ under settings $\mathrm{I}$, $\mathrm{II}$, $\mathrm{III}$, and $\mathrm{IV}$  is given as
	\begin{align*}
	\phi _{\mathrm{I}} =  \phi _{\mathrm{II}} =  \phi _{\mathrm{III}} ={} \phi_{\mathrm{IV}} =  \begin{pmatrix}
	    \dfrac{G}{1-q} \left[  \dfrac{A\{Y - \mu_1(X)\}}{e_1(X)} - \dfrac{(1-A)\{Y - \mu_0(X)\}}{1 - e_1(X)} \right] \dfrac{1 - \pi(X)}{\pi(X)} \\ 
 ~~~	+{}  \dfrac{1-G}{1-q} \{ \mu_1(X) - \mu_0(X) - \tau \}
	\end{pmatrix}. 
	\end{align*}
The associated semiparametric efficiency bound is $\mathbb{V}_{\mathrm{I}}^\ast = \mathbb{V}_{\mathrm{II}}^\ast  =\mathbb{V}_{\mathrm{III}}^\ast =\mathbb{V}_{\mathrm{IV}}^\ast  =  \E(\phi_{\mathrm{I}}^2)$.	
\item[(b)]  the efficient influence function of $\tau$ under setting $\mathrm{V}$  is given as
	\begin{align*}
	\phi_{\mathrm{V}} =\begin{pmatrix}
	    \dfrac{1 - \pi(X)}{1-q} \left [  \dfrac{GA\{Y - \mu_1(X)\}}{\tilde e(X)} -  \dfrac{(1-A)\{Y - \mu_0(X)\} }{ 1-\tilde e(X)} \right ] \\ + \dfrac{1-G}{1-q}\{\mu_1(X) -  \mu_0(X) -  \tau\}
	\end{pmatrix}, 
	\end{align*}
 where $\tilde e(X)=e_1(X)\pi(X)$. 
The semiparametric efficiency bound is $\mathbb{V}_{\mathrm{V}}^\ast =  \E(\phi_{\mathrm{V}}^2)$. 
\item[(c)] the efficient influence function of $\tau$ under setting $\mathrm{VI}$ is given as
			\begin{align*}
	\phi_{\mathrm{VI}} ={}& \frac{1 - \pi(X)}{1-q}  \left[\frac{A\{Y - \mu_1(X)\}  }{ e(X)}  -  \frac{(1-A)\{Y - \mu_0(X)\} }{  1-e (X) } \right]  +  \frac{1-G}{1-q}  \{\mu_1(X) -  \mu_0(X) -  \tau\}, 
	\end{align*}
 where $e(X) =e_0(X) \{1 - \pi(X)\} +  e_1(X)\pi(X)$. 
The semiparametric efficiency bound is $\mathbb{V}_{\mathrm{VI}}^\ast =  \E(\phi_{\mathrm{VI}}^2)$.		
In particular,  when $e_0(X) = 0$, $\phi_{\mathrm{VI}}$ reduces to $\phi_{\mathrm{V}}$;   when $\pi(X) = 0$, $\phi_{\mathrm{VI}}$ reduces to the efficient influence function based only the target data, i.e., 
     \[  \frac{1-G}{1-q}  \left[\frac{A\{Y - \mu_1(X)\}  }{ e_0(X) }  -  \frac{(1-A)\{Y - \mu_0(X)\} }{ 1-e_0(X) } + \{\mu_1(X) -  \mu_0(X) -  \tau\} \right].   \]
\end{itemize}
	
\end{theorem}

Theorem \ref{thm1}(a) states that in settings I-IV, where only  minimal assumptions (Assumptions \ref{assump1}-\ref{assump2}) are imposed on the underlying data-generating distribution, the efficient information for estimating $\tau$ is solely derived from the covariates $X$. Intuitively, given that Assumptions \ref{assump1}-\ref{assump2} allow the existence of unobserved confounders,  collecting  the additional data $(A, Y)$ does not alter the semiparametric efficiency bound of $\tau$. This suggests that the impact on estimation efficiency in these situations primarily depends on the assumptions made, rather than the data collected.  

Theorems \ref{thm1}(b)-(c) demonstrate that introducing additional assumptions, namely Assumption \ref{assump3} or Assumption \ref{assump4}, to the collected data leads to changes in the efficiency bounds of the parameter $\tau$. In these scenarios, collecting additional data $(A, Y)$ provides benefit in enhancing the estimation efficiency of $\tau$. 
In addition, as shown in Theorem \ref{thm1}(c), despite the mutual exclusivity of Assumptions \ref{assump3} and \ref{assump4} (i.e., they cannot coexist), the form of $\phi_{\mathrm{VI}}$ encompasses both $\phi_{\mathrm{V}}$ and the EIF based only on the target data in certain degenerate cases ($e_0(X) = 0$ or $\pi(X) = 0$). This is consistent with the fact that Assumption \ref{assump4} is a special case of Assumption \ref{assump3}(i), and Assumption \ref{assump3} alone is sufficient to identify $\tau$ when $\pi(X) = 0$.

Notably, the EIFs $\phi_{\mathrm{V}}$ and $\phi_{\mathrm{VI}}$ can be expressed in a unified form given by 
    			\begin{align*}
 \frac{1 - \pi(X)}{1-q} & \left[ \frac{A\{Y - \mu_1(X)\}}{ \P(A=1\mid X) }  -  \frac{(1-A)\{Y - \mu_0(X)\}}{ \P(A=0\mid X) } \right] +  \frac{1-G}{1-q} \{\mu_1(X) -  \mu_0(X) -  \tau\}. 
	\end{align*}	
This follows immediately by noting that under setting V, $\P(A=1\mid X) = \P(A=1\mid X, G=1) \P(G=1\mid X) + \P(A=1\mid X, G=0) \P(G=0\mid X) = e_1(X)\pi(X) = \tilde e(X)$, as presented in Theorem \ref{thm1}(b). Similarly, under setting VI,  $\P(A=1\mid X) = e_0(X)\{1 - \pi(X)\} + e_1(X)\pi(X) = e(X)$, as presented in Theorem \ref{thm1}(c).  

The EIFs of Settings I and V have previously been derived in \cite{Li-etal2023} and \cite{li2023improving}, respectively. 
However, the EIFs for other settings are relatively new findings, and there is insufficient comparative analysis regarding the relative efficiency among different settings.  
 Given that Assumption \ref{assump3} and Assumption \ref{assump4} cannot simultaneously hold,  
we thus introduce the notation $\tilde \P$ and $\P$ to represent the superpopulation models under settings V and VI, respectively, for comparing the two populations.

\begin{theorem}[efficiency comparison of $\tau$] \label{thm2}  The semiparametric efficiency bounds of $\tau$ under settings $\mathrm{I}$-$\mathrm{VI}$ satisfy that 
    \begin{align*}  
     \mathbb{V}_{\mathrm{I}}^\ast = \mathbb{V}_{\mathrm{II}}^\ast  =\mathbb{V}_{\mathrm{III}}^\ast =\mathbb{V}_{\mathrm{IV}}^\ast >  \mathbb{V}_{\mathrm{V}}^\ast, \quad  \mathbb{V}_{\mathrm{I}}^\ast = \mathbb{V}_{\mathrm{II}}^\ast  =\mathbb{V}_{\mathrm{III}}^\ast =\mathbb{V}_{\mathrm{IV}}^\ast  > \mathbb{V}_{\mathrm{VI}}^\ast. 
    \end{align*}
More specifically, 
\begin{itemize}
    \item[(a)] the efficiency gain of setting $\mathrm{V}$ compared to setting $\mathrm{I}$ is 
    \begin{align*}
		\mathbb{V}_{\mathrm{I}}^\ast  - \mathbb{V}_{\mathrm{V}}^\ast 
		 ={}&  \E \Biggl [  \gamma_1(X)  \frac{\{1 - \pi(X)\}^2\mathrm{Var}\{Y(0) \mid X\}}{(1-q)^2\pi(X)\{1 - e_1(X)\}}  \Biggr  ], 
	\end{align*}
  where $\gamma_1(X)  = \dfrac{ \tilde \P(A=0, G =0\mid X)  }{   \tilde \P(A=0 \mid X) } = \dfrac{ 1 - \pi(X) }{  \{1 -\pi(X)\} + \{1 - e_1(X)\}\pi(X) }.$
    
    \item[(b)] 
the efficiency gain of setting $\mathrm{VI}$ compared to setting $\mathrm{I}$  is 
     	 \begin{align*} 
	   \mathbb{V}_{\mathrm{I}}^\ast  - \mathbb{V}_{\mathrm{VI}}^\ast 
	   ={}&      \E \left( \frac{ \{1-\pi(X)\}^2 }{(1-q)^2 \pi(X)}     \Biggl [ \alpha_2(X)  \frac{\mathrm{Var}\{Y(1) \mid X\}}{e_1(X) }    +  \gamma_2(X)    \frac{ \mathrm{Var}\{Y(0) \mid X\} }{ 1 - e_1(X) }  \Biggr ] \right ) ,
	   \end{align*} 
where $ \alpha_2(X) = \dfrac{\P(A=1, G = 0\mid X)}{\P(A=1\mid X)}$ and $\gamma_2(X) = \dfrac{\P(A=0, G=0\mid X)}{\P(A=0\mid X)}.$   

    \item[(c)]  the difference between $\V_{\mathrm{V}}^\ast$ and $\V_{\mathrm{VI}}^\ast$ is  
\begin{align*}
     \mathbb{V}_{\mathrm{V}}^\ast  - \mathbb{V}_{\mathrm{VI}}^\ast ={}&   \E \left (\frac{\{1 - \pi(X)\}^2}{(1-q)^2} \left[ \alpha_3(X)  \frac{\mathrm{Var}\{Y(1) \mid X\}}{\tilde  \P(A=1\mid X) }  - \gamma_3(X) \frac{\mathrm{Var}\{Y(0) \mid X\}}{\tilde \P(A=0 \mid X)} \right] \right),
\end{align*}
where $\alpha_3(X) = \alpha_2(X)$ and  $ \gamma_3(X) =  {\P(A=1, G=0\mid X) }/{  \P(A=0 \mid X) }$.   
\end{itemize}
\end{theorem} 

Theorem \ref{thm2} indicates that there is no efficiency gain from observing $(A, Y)$ in the target data under Assumptions \ref{assump1}-\ref{assump2}.  When we further invoke Assumption \ref{assump3} or \ref{assump4}, a  efficiency gain emerges.   
Comparing setting IV with setting V is interesting, especially because of the counterintuitive result that $\V_{\mathrm{IV}}^\ast > \V_{\mathrm{V}}^\ast$. In setting IV, we have access to all covariates, treatment, and outcome. 
If we simply discard observations with $A=1$ in the target data and only use those with $A=0$, setting V may seem to reduce to setting IV and a efficiency gain may be obtained. However, this is not the case because the fact that all observations of $A=0$ in the target data does not guarantee that their propensity scores are strictly 0 (i.e., $e_0(X)=0$ if $A=0$), and the unconfoundedness assumption may still not hold in the sub-samples of $A=0$. In other words, the observations with a value of $A=0$ in the target data are generated by chance, rather than deterministically.

Theorems \ref{thm2}(a)-(c) reveal the key factors affecting the efficiency improvement from setting I to setting VI.  
 By direct algebras (see the proof of Theorem \ref{thm2}), 
 	\begin{align*}
	 \mathbb{V}_{\mathrm{I}}^\ast  =
	       \E  \left ( \frac{\{1 - \pi(X)\}^{2}}{(1-q)^2\pi(X)}  \left [ \dfrac{ \mathrm{Var}\{Y(1)\mid X\}}{e_1(X)} +  \dfrac{\mathrm{Var}\{Y(0)\mid X\}}{   1 - e_1(X) } \right]\right )  +  \frac{\E \left[ \{ \mu_1(X) -  \mu_0(X)  -  \tau \}^2 \pi(X) \right]}{(1-q)^2},  
	\end{align*} 
 which gets gradually larger as $\mathrm{Var}\{Y(1)\mid X\}$ and $\mathrm{Var}\{Y(0)\mid X\}$ increase and is affected by $\pi(X)$. 
 When comparing the efficiency bounds among settings $\mathrm{I}$-$\mathrm{VI}$, in addition to $\pi(X)$, $\mathrm{Var}\{Y(1)\mid X\}$ and $\mathrm{Var}\{Y(0)\mid X\}$,  Theorem \ref{thm2}(a) shows that the efficiency gain $\mathbb{V}_{\mathrm{I}}^\ast  - \mathbb{V}_{\mathrm{V}}^\ast$ mainly depends on $\gamma_1(X)$, i.e., the proportion of controls in the target data to the whole controls, and Theorem \ref{thm2}(b) shows that the efficiency gain $\mathbb{V}_{\mathrm{I}}^\ast  - \mathbb{V}_{\mathrm{VI}}^\ast$ mainly relies on $\alpha_2(X)$ and  $\gamma_2(X)$, i.e., the proportion of treated units in the target data to the whole treated units and  the proportion of controls in the target data to the whole controls, respectively. 
In addition, as $\pi(X)$ increases, $\gamma_1(X)$, $\alpha_2(X)$, and $\gamma_2(X)$ tend to be smaller.  
In Section \ref{sim-studies}, we will explore the effect of $\pi(X)$, $\mathrm{Var}\{Y(1)\mid X\}$ and $\mathrm{Var}\{Y(0)\mid X\}$ on the estimators of $\tau$ under settings I to VI by performing a sensitivity analysis empirically. 

  As shown in Theorems \ref{thm2}(a) and \ref{thm2}(b), setting VI enhances the efficiency of $\tau$ in both treatment and control groups compared to setting I, whereas setting V improves the efficiency only  in the control group. However, this does not necessarily imply that $\mathbb{V}_{\mathrm{VI}}^\ast$ will be smaller than $\mathbb{V}_{\mathrm{V}}^\ast$. In fact, according to Theorem \ref{thm2}(c), the difference $\mathbb{V}_{\mathrm{V}}^\ast - \mathbb{V}_{\mathrm{VI}}^\ast$ depends on the extent of improvement relative to setting I. 
  
In setting VI, $\tau$ is identifiable using only the target data. It is natural to compare the efficiency bound under setting VI with those based on target data only.   
\begin{proposition}  \label{prop1}
 Under setting $\mathrm{VI}$, if we estimate $\tau$ based only on the target data, the semiparametric efficiency bound is 
 		\begin{align*}
	 \mathbb{V}^\ast =  
	      \E  \left(  \frac{1-\pi(X)}{(1-q)^2}  \left[\dfrac{\mathrm{Var}\{Y(1)\mid X\} }{ e_0(X) } +  \dfrac{ \mathrm{Var}\{Y(0)\mid X\}}{ 1 - e_0(X) }  \right] \right)  + \frac{\E  [ \{ \mu_1(X) -  \mu_0(X)  -  \tau\}^2 \pi(X) ]}{(1-q)^2}.  
	\end{align*}
The efficiency gain from  leveraging the source data is 
	\[   \V^\ast - \V_{\mathrm{VI}}^\ast =   \E    
 \left[ 	    \dfrac{\{1-\pi(X)\}\alpha_4(X)}{(1-q)^2}  \dfrac{ \mathrm{Var}\{Y(1) \mid X\}}{e_0(X) } +  	    \dfrac{\{1-\pi(X)\}\gamma_4(X)}{(1-q)^2}  \dfrac{ \mathrm{Var}\{Y(0) \mid X\}}{1 - e_0(X) }     \right ]   ,
	\]  
where $    \alpha_4(X) = \dfrac{\P(A=1, G = 1 \mid X)}{\P(A=1\mid X)}$ and $\gamma_4(X) = \dfrac{\P(A=0, G = 1\mid X)}{\P(A=0 \mid X)}$. 
\end{proposition}

Unlike Theorem \ref{thm2}(b), Proposition \ref{prop1} shows that the efficiency gain from utilizing the source data primarily hinges on the proportion of treated units in the \emph{source data} compared to the whole treated units and the proportion of controls in the \emph{source data} compared to the whole controls. Moreover, $\alpha_4(X)$ and $\gamma_4(X)$ tends to be larger as $\pi(X)$ increases. 



Motivated by \cite{Li-etal2023} that investigated the role of sampling score $\pi(X)$ and propensity score $e_1(X)$ under setting I, we extend theirs analysis across settings $\mathrm{I}$-$\mathrm{VI}$. 

\begin{proposition}  \label{prop2} The following statements hold: 

(a) if the sampling score $\pi(X)$ is known,  all the efficient influence functions of $\tau$ in settings I to VI have the same form as those in Theorem \ref{thm1}, with the only difference being that the term
     $ (1-G)\{\mu_1(X) -  \mu_0(X) -  \tau\}/(1-q)$
is replaced by      $\{1-\pi(X)\} \{\mu_1(X) -  \mu_0(X) -  \tau\}/(1-q)$. In addition,  under any of settings I to VI, the efficiency gain resulting from knowing the sampling score is  ${\E[ \{\mu_1(X) -  \mu_0(X) -  \tau\}^2 \pi(X)\{1-\pi(X)\}   ]}/{(1-q)^2}.  $

(b) if the propensity score $e_1(X)$ in the source data is known, all the efficient influence functions of $\tau$ in settings I to VI 
remain unchanged.  
\end{proposition}

Proposition \ref{prop2} shows that having knowledge of the sampling score $\pi(X)$ enhances the efficiency of estimating $\tau$, while $e_1(X)$ is ancillary for estimating $\tau$. This finding aligns with the results presented in \cite{Li-etal2023}.  
In Section \ref{sec:ext-ATT}, we extend the efficiency comparative analysis to the ATT in the target population across the six settings.  

\section{Extended Comparison with Posterior Drift}
\label{sec:sen} 

In the previous section, we focused on Assumption \ref{assump2}, which considers only the covariate shift, allowing for differences in the marginal distributions of covariates $X$ between the two populations. In this section, we broaden the comparative analysis to include scenarios from the second column of Table \ref{tab:scenarios}, which involve both covariate shift and posterior drift. 
Let \(m_{a}(\cdot) = \psi_{a}^{\prime}(\cdot)\) denote the first-order derivative of \(\psi_{a}(\cdot)\) in Assumption \ref{assump5} for \(a=0,1\). For simplicity, we define \(r_0(X) = m_{0}^2\{\mu_{0}(X)\} r_0^\circ(X)\) and \(r_1(X) = m_{1}^2\{\mu_{1}(X)\} r_1^\circ(X)\), where  
$$r_a^\circ(X) =  { \mathrm{Var}\{Y(a) \mid X, G=1\}}/{\mathrm{Var}\{Y(a) \mid X, G=0\}}.$$  
These ratios characterize the relative conditional variances between the two populations and will be used to characterize estimation efficiency and ``effective sample size".
It is worth noting that under Assumption \ref{assump2}, both $r_0(X)$ and $r_1(X)$ are equal to 1.   Parallel to Theorem \ref{thm1}, we now present the corresponding EIFs and efficiency bounds for the new settings  $\mathrm{\mathrm{I}}^\ast$--$\mathrm{\mathrm{VI}}^\ast$ in Table \ref{tab:scenarios}.

\begin{theorem}[EIFs   with Assumption \ref{assump5}] \label{thm4} 
The following statements hold: 
\begin{itemize}
    \item[(a)] 
  the efficient influence function of $\tau$ under settings $\mathrm{\mathrm{I}}^\ast$, $\mathrm{II}^\ast$, $\mathrm{III}^\ast$, and $\mathrm{IV}^\ast$  is given as
 	\begin{align*}
	   \phi _{\mathrm{\mathrm{I}}^\ast} = \phi _{\mathrm{II}^\ast} = \phi _{\mathrm{III}^\ast} = \phi_{\mathrm{IV}^\ast}= \begin{pmatrix} 
	 \dfrac{G }{1-q}  \dfrac{1 - \pi(X)}{\pi(X)}   \dfrac{ m_{1} \{\mu_1(X)\}   A \{Y - \mu_1(X)\}}{e_1(X)} \\  \addlinespace[1mm]- \dfrac{G }{1-q}  \dfrac{1 - \pi(X)}{\pi(X)}  \dfrac{   m_{0} \{\mu_0(X)\}  (1-A) \{Y - \mu_0(X)\}}{1 - e_1(X)} \\\addlinespace[1mm]
 +{} \dfrac{1-G}{1-q}  [ {\psi_{1}\{\mu_1(X)\} - \psi_{0}\{\mu_0(X)\} }  -  \tau]
	   \end{pmatrix}.
	\end{align*}
The associated efficiency bound is $\mathbb{V}_{\mathrm{\mathrm{I}}^\ast}^\ast = \mathbb{V}_{\mathrm{II}^\ast}^\ast  =\mathbb{V}_{\mathrm{III}^\ast}^\ast =\mathbb{V}_{\mathrm{IV}^\ast}^\ast  =  \E(\phi_{\mathrm{\mathrm{I}}^\ast}^2)$.	 

    \item[(b)] the efficient influence function of $\tau$ under setting $\mathrm{V}^\ast$ is given as
 {\begin{align*}
	\phi_{\mathrm{V}^\ast} &={}   \begin{pmatrix}  \dfrac{{G}\{1 -\pi(X)\} }{1-q} 
	    \dfrac{ m_{1} \{\mu_1(X)\}   A \{Y - {\mu_1(X)}\} }{ \pi(X)e_1(X) } \\  \addlinespace[1mm]- \dfrac{{G}\{1 -\pi(X)\} }{1-q} \dfrac{  m_{0} \{\mu_0(X)\}  (1-A) \{Y - {\mu_0(X)}\}  }{  \pi(X)\{1-e_1(X)\} + \{1-\pi(X)\} r_0(X) } \\\addlinespace[1mm]-\dfrac{({1-G}) \{1 - \pi(X)\} }{1-q} \dfrac{ r_0(X) (1-A) [Y - \psi_0\{\mu_0(X)\}]  }{   \pi(X)\{1-e_1(X)\} +  \{1-\pi(X)\} r_0(X)  }\\ \addlinespace[1mm]+{}  \dfrac{1-G}{1-q}  [ {\psi_{1}\{\mu_1(X)\} - \psi_{0}\{\mu_0(X)\} }  -  \tau]
	\end{pmatrix} .
	\end{align*}}
The associated efficiency bound is $\mathbb{V}_{\mathrm{V}^\ast}^\ast =  \E(\phi_{\mathrm{V}^\ast}^2)$. 

    \item[(c)]  the efficient influence function of $\tau$ under setting $\mathrm{\mathrm{VI}}^\ast$ is given as
 			\begin{align*}
	\phi_{\mathrm{\mathrm{VI}}^\ast} &={}  \begin{pmatrix} \dfrac{{G}\{1 -\pi(X)\} }{1-q} 
	     \dfrac{ m_{1} \{\mu_1(X)\}   A \{Y - {\mu_1(X)}\} }{ \pi(X)e_1(X) + \{1-\pi(X)\}e_0(X) r_1(X) }  \\ \addlinespace[1mm]- \dfrac{{G}\{1 -\pi(X)\} }{1-q}  \dfrac{  m_{0} \{\mu_0(X)\}  (1-A) \{Y - {\mu_0(X)}\}  }{  \pi(X)\{1-e_1(X)\} + \{1-\pi(X)\}\{1-e_0(X)\} r_0(X) } 
\\\addlinespace[1mm]
   +  \dfrac{({1-G}) \{1 - \pi(X)\} }{1-q}  \dfrac{ r_1(X)  A[Y - \psi_1\{\mu_1(X)\}]  }{ \pi(X)e_1(X) + \{1-\pi(X)\}e_0(X) r_1(X)  } \\  \addlinespace[1mm]  -    \dfrac{({1-G}) \{1 - \pi(X)\} }{1-q} \dfrac{ r_0(X)  (1-A)[Y -\psi_0\{ {\mu_0(X)}\}]  }{   \pi(X)\{1-e_1(X)\} + \{1-\pi(X)\}\{1-e_0(X)\} r_0(X)  } \\ \addlinespace[1mm]+{}  \dfrac{1-G}{1-q}  \big[ {\psi_{1}\{\mu_1(X)\} - \psi_{0}\{\mu_0(X)\} }  -  \tau\big]
 \end{pmatrix}.
	\end{align*}
 The associated efficiency bound is $\mathbb{V}_{\mathrm{\mathrm{VI}}^\ast}^\ast =  \E(\phi_{\mathrm{\mathrm{VI}}^\ast}^2)$.	
\end{itemize}	
\end{theorem}


Theorem \ref{thm4} reveals the pivotal role played by $r_0(X)$ and $r_1(X)$ in the efficiency bounds when covariate shift, posterior drift, or both are present. We first observe that Assumption \ref{assump2} indicates that both the first and second moments of the potential outcomes given \(X\) are equal between the two populations, leading to  \(r_0(X) = r_1(X) = 1\) and \(m_0(u) = m_1(u) = 1\) for all $u$. Consequently, Theorem \ref{thm4} precisely simplifies to Theorem \ref{thm1}. Next, when the mean exchangeability assumption holds, i.e., \(\psi_{0}(u) = \psi_{1}(u) = u\) and \(m_0(u) = m_1(u) = 1\) for all \(u\), Assumption \ref{assump5} only considers the first-order moment condition \(\tilde \mu_a(X) = \mu_a(X)\), without imposing any restrictions on the conditional variance (i.e., the second-order moment) of potential outcomes. In this case, \(r_0(X)\) and \(r_1(X)\) reflect the variance ratio between the two populations and impact the semiparametric efficiency bounds, which is consistent with the findings of \cite{li2023improving}. Finally, in the presence of both covariate shift and posterior drift, the efficiency bounds for \(\tau\) are influenced not only by the variance ratios \(r_0(X)\) and \(r_1(X)\) but also by the first-order derivatives of $m_0(u) $ and $m_1(u) $. 

\emph{Essentially, \(r_0(X)\) and \(r_1(X)\) quantify the ``effective sample size'' ratio for the control and treated units in the target dataset, respectively. Both \(r_0(X)\) and \(r_1(X)\) can be viewed as optimal transfer rates.} These optimal transfer rates are   equal to 1 under Assumption \ref{assump2}. 
Specifically, when each control (treated) unit in the target population is weighted by \(r_0(X)\) (\(r_1(X)\)), and the weighted target population is treated as the new target population. Then, under setting   \(\mathrm{V}^\ast\), \(\tilde \P_w(A=1 \mid X) = \pi(X)e_1(X)\) and \(\tilde \P_w(A=0 \mid X) = \pi(X)\{1-e_1(X)\} + \{1-\pi(X)\} r_0(X)\);   under setting \(\mathrm{VI}^\ast\), \(\P_w(A=1 \mid X) = \pi(X)e_1(X) + \{1-\pi(X)\}e_0(X) r_1(X)\) and \(\P_w(A=0 \mid X) = \pi(X)\{1-e_1(X)\} + \{1-\pi(X)\}\{1-e_0(X)\}r_0(X)\),  where \(\tilde \P_w\) and \(\P_w\) represent the weighted superpopulation models under settings \(\mathrm{V}^\ast\) and \(\mathrm{VI}^\ast\).
\begin{theorem}[efficiency comparison   with Assumption \ref{assump5}] \label{thm5}  The semiparametric efficiency bounds of $\tau$ under settings $\mathrm{\mathrm{I}}^\ast$-$\mathrm{\mathrm{VI}}^\ast$ satisfy that 
    \begin{align*}  
     \mathbb{V}_{\mathrm{\mathrm{I}}^\ast}^\ast = \mathbb{V}_{\mathrm{II}^\ast}^\ast  =\mathbb{V}_{\mathrm{III}^\ast}^\ast =\mathbb{V}_{\mathrm{IV}^\ast}^\ast >  \mathbb{V}_{\mathrm{V}^\ast}^\ast, \quad  \mathbb{V}_{\mathrm{\mathrm{I}}^\ast}^\ast = \mathbb{V}_{\mathrm{II}^\ast}^\ast  =\mathbb{V}_{\mathrm{III}^\ast}^\ast =\mathbb{V}_{\mathrm{IV}^\ast}^\ast  > \mathbb{V}_{\mathrm{\mathrm{VI}}^\ast}^\ast. 
    \end{align*}
\begin{itemize}
    \item[(a)] the efficiency gain $\mathbb{V}_{\mathrm{\mathrm{I}}^\ast}^\ast  - \mathbb{V}_{\mathrm{V}^\ast}^\ast $ between settings $\mathrm{\mathrm{I}}^\ast$-$\mathrm{\mathrm{VI}}^\ast$ and $\mathrm{V}^\ast$ is
    \begin{align*}
		\mathbb{V}_{\mathrm{\mathrm{I}}^\ast}^\ast  - \mathbb{V}_{\mathrm{V}^\ast}^\ast 
		 ={}&  \E \Biggl [   \frac{\{1 - \pi(X)\}^2}{(1-q)^2 \pi(X)} \tilde \gamma_1(X) \frac{m_{0} ^2\{\mu_0(X)\}   \mathrm{Var}\{Y(0) \mid X, G=1\}}{\{1 - e_1(X)\}}  \Biggr  ], 
	\end{align*}
  where   $\tilde \gamma_1(X)  =  \tilde \P_w(A=0, G =0\mid X)/\tilde \P_w(A=0 \mid X)$.  
    \item[(b)] the efficiency gain $\mathbb{V}_{\mathrm{\mathrm{I}}^\ast}^\ast  - \mathbb{V}_{\mathrm{\mathrm{VI}}^\ast}^\ast $ between settings $\mathrm{\mathrm{I}}^\ast$-$\mathrm{\mathrm{VI}}^\ast$ and $\mathrm{\mathrm{VI}}^\ast$ is
  \begin{align*}  
    \E    \left (\dfrac{ \{1-\pi(X)\}^2  }{ (1-q)^2 \pi(X)} 
	  \begin{bmatrix}
	      \tilde \alpha_2(X)  \dfrac{m_{1} ^2\{\mu_1(X)\} \mathrm{Var}\{Y(1)\mid X, G=1\}}{e_1(X) }\\
       \addlinespace[1mm]+  \tilde \gamma_2(X)  \dfrac{ m_{0} ^2\{\mu_0(X)\} \mathrm{Var}\{Y(0)\mid X, G=1\} }{ 1 - e_1(X)}
	  \end{bmatrix} \right ) 
	   \end{align*} 
where $\tilde \alpha_2(X)   = \P_w(A=1, G = 0\mid X)/\P_w(A=1\mid X)$ and $\tilde \gamma_2(X)  =\P_w(A=0, G=0\mid X)/\P_w(A=0\mid X)$.  
\item[(c)] the difference $\mathbb{V}_{\mathrm{V}^\ast}^\ast  - \mathbb{V}_{\mathrm{\mathrm{VI}}^\ast}^\ast$ is 
\begin{align*}
  \E \left (  \dfrac{\{1-\pi(X)\}^2 }{(1-q)^2} \left [ \tilde \alpha_3(X)   \dfrac{r_1(X)\mathrm{Var}\{Y(1)\mid X, G=0\}}{\tilde \P_w(A=1\mid X) } -  \tilde \gamma_3(X) \dfrac{ r_0(X) \mathrm{Var}\{Y(0)\mid X, G=0\}}{\tilde \P_w(A=0 \mid X)} \right ]\right ),
\end{align*}
where $ \tilde \alpha_3(X) = \tilde \alpha_2(X)$ and $\tilde \gamma_3(X) = \P_w(A=1, G=0\mid X)/\P_w(A=0 \mid X)$.  
\end{itemize} 
\end{theorem}

From Theorems \ref{thm5}(a) and \ref{thm5}(b), one can see that introducing additional unconfounded dataset always reduces the efficiency bounds of $\tau$ for any given drift functions $\psi_{0}(\cdot)$ and $\psi_{1}(\cdot)$.
Similar to Proposition \ref{prop1}, the efficiency bound in setting $\mathrm{\mathrm{VI}}^\ast$ is lower than that based solely on the target data, as shown in Proposition \ref{prop3}.  
\begin{proposition}  \label{prop3}
 Under setting $\mathrm{\mathrm{VI}}^\ast$, if we estimate $\tau$ based only on the target data and the semiparametric efficiency bound is denoted by $\mathbb{V}^\ast$. 
Then efficiency gain from  leveraging the source data is given as 
	\[   \V^\ast - \V_{\mathrm{\mathrm{VI}}^\ast}^\ast =    \E   \begin{pmatrix}
	  \dfrac{1-\pi(X)}{(1-q)^2} \left[  \dfrac{ \tilde \alpha_4(X)\mathrm{Var}\{Y(1) \mid X, G=0\}}{e_0(X) } +   \dfrac{\tilde \gamma_4(X) \mathrm{Var}\{Y(0) \mid X, G=0\}}{1 - e_0(X) }     \right ]  
	\end{pmatrix}, 
	\]  
where $\tilde \alpha_4(X) =  \dfrac{\P_w(A=1, G = 1 \mid X)}{\P_w(A=1\mid X)}$ and $ \tilde \gamma_4(X) = \dfrac{\P_w(A=0, G = 1\mid X)}{\P_w(A=0 \mid X)}$. 
\end{proposition}

\section{Estimation and Inference} \label{sec-estimation}
In this section, we propose efficient estimators for $\tau$   under various settings in Table \ref{tab:scenarios}. 

\subsection{Estimation and Inference under Covariate Shift}
\label{sec5}
We first propose efficient estimators for $\tau$ under Settings I-IV and present the large sample properties of the proposed estimators.  For simplicity, we denote the empirical mean operator by $\mathbb{P}_n$, defined as $\P_n(U) = n^{-1}\sum_{i=1}^n U_i$, where $U$ represents a generic variable. 
Let $\hat{\pi}(X)$ be a fitted working model for the sampling score $\pi(X)$, $\hat{e}_a(X)$ a fitted working model for the propensity score $e_a(X)$ for $a = 0, 1$, and $\hat{\mu}_a(X)$ a fitted working model for the outcome model $\mu_a(X)$ for $a = 0, 1$.  
In practice, many methods can be used to estimate the nuisance parameters. For example, we can specify parametric models and consider maximum likelihood estimation or the method of moments. Additionally, nonparametric methods, such as random forests and sieve methods \citep{chen2007large}, combined with cross-fitting techniques~\citep{chernozhukov2018double}, can also be used to estimate nuisance parameters. 

For Settings I-IV, we use the datasets $(G=0, A=0)$ and $(G=0, A=1)$ to estimate $\mu_0(X)$ and $\mu_1(X)$, respectively. In Setting VI, $\mu_0(X)$ and $\mu_1(X)$ are estimated in a similar manner. However, to fully utilize the observed data under Assumption \ref{assump3} or \ref{assump4}, we also recommend combining the source and target datasets to estimate $\mu_0(X)$ and $\mu_1(X)$ using the datasets where $(A=0)$ and $(A=1)$, respectively.  Theorem \ref{thm1} motivates the following plug-in estimators for settings $\mathrm{I}$-$\mathrm{VI}$: 
  \begin{equation*} 
        \begin{gathered}  
\hat \tau_{\mathrm{I}} = \hat \tau_{\mathrm{II}} = \hat \tau_{\mathrm{III}} = \hat \tau_{\mathrm{IV}}={}  \P_n \begin{pmatrix}
     \dfrac{G}{1-q} \left [   \dfrac{A\{Y - \hat \mu_1(X)\}}{\hat e_1(X)} - \dfrac{(1-A)\{Y - \hat \mu_0(X)\}}{1 - \hat e_1(X)} \right]\dfrac{1 - \hat \pi(X)}{\hat \pi(X)}\\\addlinespace[1mm] + \dfrac{1-G}{1-q} \{ \hat \mu_1(X) - \hat \mu_0(X) \}
\end{pmatrix}, \\\addlinespace[1.5mm]
\hat \tau_{\mathrm{VI}} ={}  \P_n \begin{pmatrix}\dfrac{ 1 - \hat \pi(X) }{1-q} \left[ \dfrac{A\{Y - \hat \mu_1(X)\}  }{\hat e( X)} - \dfrac{(1-A)\{Y - \hat \mu_0(X)\}}{1-\hat e(X)} \right ]\\\addlinespace[1mm] + \dfrac{1-G}{1-q}\{\hat \mu_1(X) - \hat \mu_0(X) \} \end{pmatrix},   
\end{gathered}  
\end{equation*}
where $\hat e( X) = \hat e_0(X)\{1 - \hat \pi(X)\} + \hat \pi(X) \hat e_1(X)$.  
Based on Theorem \ref{thm1}, we know that $\hat{\tau}_{\mathrm{V}}$ follows the same form as $\hat{\tau}_{\mathrm{VI}}$, except that $\hat{e}_0(X)$ in $\hat{e}(X)$ is replaced by 0. The following Theorem \ref{thm3-old} demonstrates the double robustness of the proposed estimators.

\begin{theorem} 
\label{thm3-old}
The following statements hold:
\begin{enumerate}
    \item[(a)] Given the assumptions in Setting $\mathrm{I}$, $\hat{\tau}_{\mathrm{I}}$ is consistent if either of the following conditions is satisfied: (i) $\hat{\mu}_a(X) \xrightarrow{p} \mu_a(X)$ for $a = 0, 1$, or (ii) $\hat{e}_1(X) \xrightarrow{p} e_1(X)$ and $\hat{\pi}(X) \xrightarrow{p} \pi(X)$.
    \item[(b)] Given the assumptions in Setting $\mathrm{V}$, $\hat{\tau}_{\mathrm{V}}$ is consistent if either of the following conditions is satisfied: (i) $\hat{\mu}_a(X) \xrightarrow{p} \mu_a(X)$ for $a = 0, 1$, or (ii) $\hat{e}_1(X) \xrightarrow{p} e_1(X)$ and $\hat{\pi}(X) \xrightarrow{p} \pi(X)$.
    \item[(c)] Given the assumptions in Setting $\mathrm{VI}$, $\hat{\tau}_{\mathrm{VI}}$ is consistent if either (i) $\hat{\mu}_a(X) \xrightarrow{p} \mu_a(X)$ for $a = 0, 1$, or (ii) $\hat{e}_a(X) \xrightarrow{p} e_a(X)$ for $a = 0, 1$, and $\hat{\pi}(X) \xrightarrow{p} \pi(X)$. 
\end{enumerate}    
\end{theorem}


We now present the rate constraints for the employed working models, allowing for various estimation methods. By incorporating flexible approaches to estimate the nuisance parameters, we can reduce the risk of parametric model misspecification in practice. Under certain regularity conditions \citep{kennedy2023towards}, these nonparametric methods offer an alternative path to achieving consistency and asymptotic normality. This concept parallels the doubly debiased machine learning approach used for estimating the ATE \citep{chernozhukov2018double}. We impose the following rate restrictions on the estimators of the nuisance parameters. 
  \begin{condition}\label{cond1-old} For $a = 0, 1$,   
\begin{itemize}
    \item[(i)]$|| \hat e_1(X) - e_1(X)  ||_2   || \hat \mu_a (X)- \mu_a (X)||_2   = o_{\P}(n^{-1/2})$, \\ $|| \hat \pi(X) - \pi (X)||_2   || \hat \mu_a(X) - \mu_a(X) ||_2 = o_{\P}(n^{-1/2})$; 
 \item[(ii)]   $|| \hat e_0 (X)- e_0(X)  ||_2   || \hat \mu_a(X) - \mu_a(X) ||_2   = o_{\P}(n^{-1/2})$. 
\end{itemize}
\end{condition}  
Condition \ref{cond1-old} is a high-level requirement regarding the convergence rates of the estimators for the nuisance parameters. Condition \ref{cond1-old}(i) specifies the rate required for the working models used in Settings I-V, while Condition \ref{cond1-old}(ii) governs the rate for the working models in Setting VI.  
Condition \ref{cond1-old} is satisfied if either correctly specified parametric models are used, or flexible nonparametric methods are employed to estimate the nuisance parameters, provided that these estimation methods achieve a convergence rate faster than \(n^{-1/4}\).

\begin{theorem} \label{thm3-3} The following statements hold:
\begin{itemize}
    \item[(a)] 
Given the assumptions in Setting $\mathrm{I}$ and Condition \ref{cond1-old}(i),  
 $\sqrt{n}(\hat \tau_{\mathrm{I}} - \tau ) \xrightarrow{d} N(0,    \V_{\mathrm{I}})$, 
  where $\V_{\mathrm{I}}$ is the efficiency bound of $\tau$ under setting $\mathrm{I}$.  
    \item[(b)] 
Given the assumptions in Setting $\mathrm{V}$ and Condition \ref{cond1-old}(i),  
 $\sqrt{n}(\hat \tau_{\mathrm{V}} - \tau ) \xrightarrow{d} N(0,    \V_{\mathrm{V}})$, 
  where $\V_{\mathrm{V}}$ is the efficiency bound of $\tau$ under setting $\mathrm{V}$.   
    \item[(c)]   Given the assumptions in Setting $\mathrm{VI}$ and Conditions \ref{cond1-old}(i)--(ii),   $\sqrt{n}(\hat \tau_{\mathrm{VI}} - \tau ) \xrightarrow{d} N(0,    \V_{\mathrm{VI}}),$ 
where $\V_{\mathrm{VI}}$ is the efficiency bound of $\tau$ under setting  $\mathrm{VI}$.    
\end{itemize}
\end{theorem}

Theorem \ref{thm3-3} establishes that the proposed estimators are consistent, asymptotically normal, and locally efficient under Condition \ref{cond1-old}. According to Theorem \ref{thm3-3}, the asymptotic variances of \(\hat{\tau}_{\mathrm{I}}\), \(\hat{\tau}_{\mathrm{V}}\), and \(\hat{\tau}_{\mathrm{VI}}\) can be naturally estimated by \(\P_n (\hat{\phi}_{\mathrm{I}}^2)\), \(\P_n (\hat{\phi}_{\mathrm{V}}^2)\), and \(\P_n(\hat{\phi}_{\mathrm{VI}}^2)\), respectively, where \(\hat{\phi}_{\mathrm{I}}\), \(\hat{\phi}_{\mathrm{V}}\), and \(\hat{\phi}_{\mathrm{VI}}\) are the plug-in estimators of \(\phi_{\mathrm{I}}\), \(\phi_{\mathrm{V}}\), and \(\phi_{\mathrm{VI}}\) in Theorem \ref{thm1}. For more general cases, we recommend using the nonparametric bootstrap.

\subsection{Estimation and Inference  under both Covariate Shift and Posterior Drift} \label{est-postdrift}
In this subsection, we propose efficient estimators for \(\tau\) under Settings \(\mathrm{I}^\ast - \mathrm{IV}^\ast\). To simplify the exposition, we introduce two baseline variance ratios: 
$$
r^\circ_0(X) := \dfrac{\text{Var}(Y \mid X, A=0, G=1)}{\text{Var}(Y \mid X, A=0, G=0)}, \quad r^\circ_1(X) := \dfrac{\text{Var}(Y \mid X, A=1, G=1)}{\text{Var}(Y \mid X, A=1, G=0)},
$$
and under Assumption \ref{assump1}, we have \(r_a(X) = m_a^2(X) r_a^\circ(X)\) for \(a = 0, 1\).

Let \(\eta(X) := \{e_0(X), e_1(X), \pi(X), \mu_0(X), \mu_1(X), r_0^\circ(X), r_1^\circ(X)\}\) represent the nuisance parameters, and let \(\hat{\eta}(X) = \{\hat{e}_0(X), \hat{e}_1(X), \hat{\pi}(X), \hat{\mu}_0(X), \hat{\mu}_1(X), \hat{r}^\circ_0(X), \hat{r}^\circ_1(X)\}\) denote their estimates. Similar to the estimation method described in Section \ref{sec5}, \(\hat{\eta}(X)\) can be estimated using either a standard parametric model or a flexible nonparametric approach. 
For estimating \(r_a^\circ(X)\), one can use plug-in estimators for \(\mathrm{Var}(Y \mid X, A=a, G=1)\) and \(\mathrm{Var}(Y \mid X, A=a, G=0)\) as outlined in \citep{Fan-Yao-1998}. For example, the estimate of \(\mathrm{Var}(Y \mid X, A=a, G=1)\) can be obtained by performing a nonparametric regression of \(\{Y - \hat{\mu}_a(X)\}^2\) on \(X\) based on data with \((A=a, G=1)\). 
We then construct the estimators of \(\tau\) under Settings \(\mathrm{I}^\ast - \mathrm{IV}^\ast\) by substituting the estimated nuisance parameters \(\hat{\eta}(X)\) into the efficient influence functions (EIFs) provided in Theorem \ref{thm4}, as follows:
 
\begin{gather*} 
\hat \tau_{\mathrm{I}^\ast} =	  \hat \tau_{\mathrm{II}^\ast} =\hat \tau _{\mathrm{III}^\ast} =\hat \tau_{\mathrm{IV}^\ast}= \begin{pmatrix} 
	 \dfrac{G }{1-q}  \dfrac{1 - \hat  \pi(X)}{ \hat \pi(X)}   \dfrac{ m_{1} \{ \hat \mu_1(X)\}   A \{Y -  \hat \mu_1(X)\}}{ \hat e_1(X)} \\  \addlinespace[1mm]- \dfrac{G }{1-q}  \dfrac{1 -  \hat \pi(X)}{ \hat \pi(X)}  \dfrac{   m_{0} \{  \hat \mu_0(X)\}  (1-A) \{Y -  \hat \mu_0(X)\}}{1 -  \hat e_1(X)} \\\addlinespace[1mm]
 +{} \dfrac{1-G}{1-q}  [ {\psi_{1}\{ \hat \mu_1(X)\} - \psi_{0}\{ \hat \mu_0(X)\} }  ]
	   \end{pmatrix} , \\ 
\hat \tau_{\mathrm{V}^\ast}  ={}  \P_n  \begin{pmatrix}  \dfrac{{G}\{1 -\hat \pi(X)\} }{1-q} 
	    \dfrac{ m_{1} \{\hat \mu_1(X)\}   A \{Y - {\hat \mu_1(X)}\} }{ \hat \pi(X)\hat e_1(X) } \\  \addlinespace[1mm]- \dfrac{{G}\{1 -\hat \pi(X)\} }{1-q} \dfrac{  m_{0} \{\hat \mu_0(X)\}  (1-A) \{Y - {\hat \mu_0(X)}\}  }{ \hat  \pi(X)\{1-\hat e_1(X)\} + \{1-\hat \pi(X)\} \hat r_0(X) } \\\addlinespace[1mm]-\dfrac{({1-G}) \{1 - \hat \pi(X)\} }{1-q} \dfrac{ \hat r_0(X) { (1-A)} [Y - \psi_0\{\hat \mu_0(X)\}]  }{  \hat  \pi(X)\{1-\hat e_1(X)\} +  \{1-\hat \pi(X)\}\hat  r_0(X)  }\\ \addlinespace[1mm]+{}  \dfrac{1-G}{1-q}  [ {\psi_{1}\{\hat \mu_1(X)\} - \psi_{0}\{\hat \mu_0(X)\} }   ]
	\end{pmatrix} ,
 \\
  \hat \tau_{\mathrm{VI}^\ast}  ={}  \P_n  \begin{pmatrix}  \dfrac{{G}\{1 - \hat \pi(X)\} }{1-q} \left [\dfrac{ m_{1}\{\hat \mu_1(X)\} A \{Y - \hat \mu_1(X)\}}{ \hat \omega_1(X) }  - \dfrac{  m_{0}\{\hat \mu_0(X)\}  (1-A) \{Y - \hat \mu_0(X)\}  }{  \hat \omega_0(X) }  \right ]\\+ \dfrac{({1-G}) \{1 - \hat \pi(X)\} }{1-q} \dfrac{ \hat r_1(X)\cdot A[Y - \psi_1\{\hat \mu_1(X)\} ] }{ \hat \omega_1(X)  }  \\ -\dfrac{({1-G}) \{1 - \hat \pi(X)\} }{1-q}  \dfrac{ \hat r_0(X) \cdot (1-A)[Y - \psi_0\{\hat \mu_0(X)\}]  }{   \hat \omega_0(X)  }   \\+ \dfrac{1-G}{1-q} \big[ \psi_1\{\hat \mu_1(X)\} - \psi_0\{\hat \mu_0(X)\} \big]  \end{pmatrix} ,
  \end{gather*}
where $\hat{\omega}_1(X)=\hat{\pi}(X) \hat{e}_1(X)+\{1-\hat{\pi}(X)\} \hat{e}_0(X) \hat{r}_1(X)$ and $\hat{\omega}_0(X)=\hat{\pi}(X)\{1-\hat{e}_1(X)\}+\{1- \hat{\pi}(X)\}\{1-\hat{e}_0(X)\} \hat{r}_0(X), \hat{r}_a(X)=m_a^2\left\{\hat{\mu}_a(X)\right\} \hat{{r}}^\circ_a(X)$ for $a=0,1$. Additionally, the estimators \(\{\hat{\tau}_{\mathrm{I}}^\ast, \ldots, \hat{\tau}_{\mathrm{VI}}^\ast\}\) take the same form as \(\{\hat{\tau}_{\mathrm{I}}, \ldots, \hat{\tau}_{\mathrm{VI}}\}\)  when \(m_0(u) = m_1(u) \equiv 1\) for any \(u\) and \(r_0^\circ(X) = r_1^\circ(X) \equiv 1\).  The following proposition establishes the consistency of the proposed estimators.
\begin{proposition} \label{prop-tmp} 
The following statements hold:
\begin{enumerate}
    \item[(a)] Given the assumptions in Setting \(\mathrm{I}^*\), \(\hat{\tau}_{\mathrm{I}^*}\) is consistent and asymptotically normal if \(\hat{\mu}_a(X) \xrightarrow{p} \mu_a(X)\) for \(a = 0, 1\), regardless of whether the other working models converge in probability to the true models. Additionally, if \(\psi_1(u) = \psi_0(u) = \epsilon u\) for some known \(\epsilon\), \(\hat{\tau}_{\mathrm{I}^*}\) remains consistent if either of the following conditions is satisfied: (i) \(\hat{\mu}_a(X) \xrightarrow{p} \mu_a(X)\) for \(a = 0, 1\), or (ii) \(\hat{e}_1(X) \xrightarrow{p} e_1(X)\) and \(\hat{\pi}(X) \xrightarrow{p} \pi(X)\).
    
    \item[(b)] Given the assumptions in Setting \(\mathrm{V}^*\), \(\hat{\tau}_{\mathrm{V}^*}\) is consistent and asymptotically normal if \(\hat{\mu}_a(X) \xrightarrow{p} \mu_a(X)\) for \(a = 0, 1\), regardless of whether the other working models converge in probability to the true models. Additionally, if \(\psi_1(u) = \psi_0(u) = \epsilon u\) for some known \(\epsilon\), \(\hat{\tau}_{\mathrm{V}^*}\) remains consistent if either of the following conditions is satisfied: (i) \(\hat{\mu}_a(X) \xrightarrow{p} \mu_a(X)\) for \(a = 0, 1\), or (ii) \(\hat{e}_1(X) \xrightarrow{p} e_1(X)\) and \(\hat{\pi}(X) \xrightarrow{p} \pi(X)\).
    
    \item[(c)] Given the assumptions in Setting \(\mathrm{VI}^*\), \(\hat{\tau}_{\mathrm{VI}^*}\) is consistent and asymptotically normal if \(\hat{\mu}_a(X) \xrightarrow{p} \mu_a(X)\) for \(a = 0, 1\), regardless of whether the other working models converge in probability to the true models. Additionally, if \(\psi_1(u) = \psi_0(u) = \epsilon u\) for some known \(\epsilon\), \(\hat{\tau}_{\mathrm{VI}^*}\) remains consistent if either of the following conditions is satisfied: (i) \(\hat{\mu}_a(X) \xrightarrow{p} \mu_a(X)\) for \(a = 0, 1\), or (ii) \(\hat{\pi}(X) \xrightarrow{p} \pi(X)\) and \(\hat{e}_a(X) \xrightarrow{p} e_a(X)\) for \(a = 0, 1\).
\end{enumerate}
\end{proposition}



 As long as the outcome models \(\hat{\mu}_a(X)\) converge to the true models \(\mu_a(X)\) for \(a = 0, 1\), the estimators \(\hat{\tau}_{\mathrm{I}^*}\), \(\hat{\tau}_{\mathrm{V}^*}\), and \(\hat{\tau}_{\mathrm{VI}^*}\) are consistent. 
 Additionally, if the posterior drift functions satisfy a specific linear relationship, such that \(\psi_1(u) = \psi_0(u) = \epsilon u\) (where \(\epsilon\) is a known constant), the proposed estimators, similar to Theorem \ref{thm3-old} in Section \ref{sec5}, still exhibit double robustness. To ensure the asymptotic normality of the proposed estimators, we introduce the following conditions. 
\begin{condition}\label{cond1} For $a= 0, 1$,  
\begin{itemize}
    \item[(i)] $|| \hat e_1 (X) - e_1(X)   ||_2   || \hat \mu_a(X)  - \mu_a(X)  ||_2   = o_{\P}(n^{-1/2})$, \\
    $|| \hat \pi(X)  - \pi (X) ||_2  || \hat \mu_a(X)  - \mu_a (X) ||_2 = o_{\P}(n^{-1/2})$;  
    \item[(ii)]  $|| \hat \mu_a (X) - \mu_a(X)  ||_2^2 = o_{\P}(n^{-1/2})$;
 \item[(iii)]   $|| \hat { r}_a^\circ(X)  - {  r}_a^\circ(X)   ||_2  || \hat \mu_a (X) - \mu_a(X)  ||_2   = o_{\P}(n^{-1/2})$; 
  \item[(iv)] $|| \hat e_0 (X) - e_0 (X)  ||_2   || \hat \mu_a(X)  - \mu_a(X)  ||_2 = o_{\P}(n^{-1/2})$.
\end{itemize}
\end{condition}

Condition \ref{cond1} is similar to Condition \ref{cond1-old} and is also a high-level condition involving the convergence rates of the nuisance parameters. Conditions \ref{cond1}(i) and (iv) is identical to Condition \ref{cond1-old}, but Conditions \ref{cond1}(ii) and (iii) are new. The same as Condition \ref{cond1-old}, Condition \ref{cond1} also allows not only the use of correctly specified parametric models but also flexible nonparametric methods to estimate the nuisance parameters.
Below, we present the asymptotic properties of the proposed estimators.

\begin{theorem} \label{thm3} The following statements hold:
\begin{itemize}
    \item[(a)] 
 Given the assumptions in Setting  $\mathrm{I}^*$ and Conditions \ref{cond1}(i) and (ii), 
 $\sqrt{n}(\hat \tau_{\mathrm{I}^*} - \tau ) \xrightarrow{d} N(0,    \V_{\mathrm{I}^*}^*)$, 
  where $\V_{\mathrm{I}^*}^*$ is the efficiency bound of $\tau$ under setting $\mathrm{I}^*$. 
Additionally, if $\psi_1(u) = \psi_0(u) = \epsilon u$ for some known $\epsilon$,   Conditions \ref{cond1}(i) and (ii) can be relaxed into Condition \ref{cond1}(i). 
    \item[(b)] Given the assumptions in Setting  $\mathrm{V}^\ast$ and Conditions \ref{cond1}(i)-(iii), $\sqrt{n}(\hat \tau_{\mathrm{V}^\ast} - \tau ) \xrightarrow{d} N(0,    \V_{\mathrm{V}^\ast}^*),$ 
where $\V_{\mathrm{V}^\ast}^*$ is the efficiency bound of $\tau$ under setting  $\mathrm{V}^\ast$.  Additionally, if $\psi_1(u) = \psi_0(u) = \epsilon u$ for some known $\epsilon$,  Conditions \ref{cond1}(i)-(iii) can be relaxed into Conditions \ref{cond1}(i) and (iii). 
  
    \item[(c)]   Given the assumptions in Setting  $\mathrm{VI}^\ast$ and Conditions \ref{cond1}(i)-(iv), $\sqrt{n}(\hat \tau_{\mathrm{VI}^\ast} - \tau ) \xrightarrow{d} N(0,    \V_{\mathrm{VI}^\ast}^*),$ 
where $\V_{\mathrm{VI}^\ast}^*$ is the efficiency bound of $\tau$ under setting  $\mathrm{VI}^\ast$.  Additionally, if $\psi_1(u) = \psi_0(u) = \epsilon u$ for some known $\epsilon$,   Conditions \ref{cond1}(i)-(iv) can be relaxed into Conditions \ref{cond1}(i), (iii), (iv).  
\end{itemize}
\end{theorem}

Combining Proposition \ref{prop-tmp} and Theorem \ref{thm3}, we find that while \(r_0^\circ(X)\) and \(r_1^\circ(X)\) influence the estimation efficiency, they do not impact its consistency. A similar observation is discussed in \cite{li2023improving} when \(\psi_1(u) = \psi_0(u) = u\). Theorem \ref{thm3} establishes that the proposed estimators are consistent, asymptotically normal, and locally efficient under Assumption \ref{cond1}. Moreover, the asymptotic variances of \(\hat{\tau}_{\mathrm{I}^*}\) and \(\hat{\tau}_{\mathrm{VI}^*}\) can be naturally estimated by \(\P_n (\hat{\phi}_{\mathrm{I}^*}^2)\) and \(\P_n(\hat{\phi}_{\mathrm{VI}^*}^2)\), respectively, where \(\hat{\phi}_{\mathrm{I}^*}\) and \(\hat{\phi}_{\mathrm{VI}^*}\) are the plug-in estimators of \(\phi_{\mathrm{I}^*}\) and \(\phi_{\mathrm{VI}^*}\). For more general cases, we recommend using the nonparametric bootstrap.

\section{Sensitivity Analysis}
\label{sec:sen2}
In the previous sections, Assumptions \ref{assump2} and \ref{assump5} serve as sufficient conditions for nonparametric identification and are core assumptions for the subsequent consideration of semiparametric estimation. These assumptions are analogous to the ignorability assumption in causal inference for observational studies \citep{rosenbaum1983assessing} and the sequential ignorability assumption in mediation analysis \citep{vanderweele2015explanation}. Generally, the more baseline covariates $X$ we observe, the more plausible these assumptions become. However, in practice, it may not be feasible to collect enough covariates to completely eliminate the ``confounding" between the selection indicator $G$ and the outcome variable, which raises the need to further analyze the transferability or generalizability between the two datasets. At this point, sensitivity analysis becomes crucial in addressing this issue in the uncertainty of the transferability assumptions    \ref{assump2} and \ref{assump5}.

Although sensitivity analysis has a long history in observational studies \citep{rosenbaum1983assessing,rosenbaum2002overt}, there is relatively limited research on sensitivity analysis in the context of data fusion, apart from bias functions in data fusion \citep{dahabreh2023sensitivity} and some modeling assumptions addressing unobserved confounders \citep{nguyen2017sensitivity,nie2021covariate}. In practice, when it is unclear which setting in Table \ref{tab:6settings} is more appropriate, we recommend applying the theoretical results from Section \ref{sec:sen} for sensitivity analysis, providing a more general framework for data fusion.



We propose using a series of sensitivity functions for sensitivity analysis, where each sensitivity function can be viewed as the posterior drift function in Assumption \ref{assump5}. For example, for a continuous outcome, we can select a range of posterior drift functions such as $\mathcal{M}_0 = \{\psi_{0,\epsilon_0}(u) : \psi_{0,\epsilon_0}(u) = \epsilon_0 u\}$ and $\mathcal{M}_1 = \{\psi_{1,\epsilon_1}(u) : \psi_{1,\epsilon_1}(u) = \epsilon_1 u\}$. In this context, $\epsilon_0$ and $\epsilon_1$ are two sensitivity parameters that index the functions $\psi_{0,\epsilon_0}(u)$ and $\psi_{1,\epsilon_1}(u)$, capturing the posterior drift deviations for potential outcomes under treatment and control, respectively. Alternatively, in such linear model settings,  parameters $\epsilon_0$ and $\epsilon_1$ can be expressed as the following two ratios: 
\begin{equation}
    \label{eq:sensitivity-pars}
    \epsilon_0 :=\dfrac{\tilde{\mu}_0(X)}{{\mu}_0(X)}= \frac{\E\{Y(0) \mid X, G=0\}}{\E\{Y(0) \mid X, G=1\}},
~~\epsilon_1 :=\dfrac{\tilde{\mu}_1(X)}{{\mu}_1(X)}= \frac{\E\{Y(1) \mid X, G=0\}}{\E\{Y(1) \mid X, G=1\}}.
\end{equation}
Therefore, the sensitivity parameters \(\epsilon_0\) and \(\epsilon_1\) correspond to the two posterior drift functions defined in Assumption \ref{assump5}. The method proposed in Section \ref{sec:sen} introduces an  EIF  for each pair of posterior drift functions \(\psi_{0,\epsilon_0}(u)\) and \(\psi_{1,\epsilon_1}(u)\). Furthermore, based on the estimation approach described in Section \ref{est-postdrift}, we can construct doubly robust machine learning estimators for $\tau$ with posterior drift functions  \(\psi_{0,\epsilon_0}(u)\) and \(\psi_{1,\epsilon_1}(u)\). Under certain rate conditions on the working models, as required by Condition \ref{cond1}, these estimators can achieve the semiparametric efficiency bound. In summary, for the sensitivity function pair \(\psi_{0,\epsilon_0}(u)\) and \(\psi_{1,\epsilon_1}(u)\), we can compute point estimates and construct 95\% confidence intervals across six different settings, enabling an accurate evaluation of the range of ATEs in the target population.

It is worth noting that, within the framework of linear models, the parameters \(\epsilon_0\) and \(\epsilon_1\) act as two sensitivity parameters in our proposed sensitivity analysis framework. In practice, we only need to specify a range for \(\epsilon_0\) and \(\epsilon_1\), and by adjusting their values, we can estimate the range of \(\tau\). Furthermore, in Setting VI$^\ast$ of Table \ref{tab:6settings}, when both datasets satisfy the ignorability assumption, \(\epsilon\) becomes theoretically identifiable under certain posterior drift function classes, such as the linear function class. However, since \(X\) may be continuous or high-dimensional in practice, we still recommend specifying a range for these parameters.

In a closely related topic, \citet{dahabreh2023sensitivity} also addresses sensitivity analysis for the ATE in the target population. Their method focuses on characterizing the bias functions of two conditional expectations, specifically using the sensitivity function \(\epsilon(a, X) := \tilde{\mu}_a(X) - \mu_a(X)\) to represent the extent of posterior drift. In contrast, within the framework of linear posterior drift functions, our sensitivity analysis approach emphasizes the ratio forms presented in \eqref{eq:sensitivity-pars}, thereby complementing existing sensitivity analysis methods in the field of data fusion. 

\section{Simulation}
\label{sim-studies}
We conduct extensive simulations to evaluate the finite-sample performance of the estimators \(\tau_{\mathrm{I}}\), \(\tau_{\mathrm{V}}\), and \(\tau_{\mathrm{VI}}\) constructed using the EIFs from Theorem \ref{thm1}. The specific forms of these estimators are detailed in Section \ref{sec5}. We assess their performance across four studies: analyzing \(\tau\) under different scenarios, exploring the efficiency gain through various sampling scores, variances of potential outcomes, and model misspecification. In the first three studies, we use parametric models to estimate nuisance parameters, including linear models for outcome regressions and logistic regression for sampling and propensity scores.
Throughout this simulation, the covariates $X = (X_1, X_2)^{\intercal} \sim N(0, I_2)$, where $I_2$ is an identity matrix, and the sample size $n$ is set to 500, 1000, and 2000.   


\medskip 
\noindent 
{\bf Study I.}  
We explore the double robustness of the estimators   under various settings and consider three data-generating scenarios for $\{G, A, Y(1), Y(0)\}$. 
\begin{itemize}
	\item[\bf (C1)]   $\P(G=1\mid X) = \text{expit}\{(X_1+X_2)/2\}$, $\P(A=1\mid X, G) = \text{expit}\{(X_1 - X_2)/2 + G\}$, $Y(1) = 3 + 2 X_1 + X_2 + \epsilon(1)$, $Y(0) = 1 - 2 X_1 + 3 X_2 + \epsilon(0)$,  where $\text{expit}(x) = \exp(x)/\{1 + \exp(x)\}$,   $\epsilon(1)\sim N(0, 4)$ and $\epsilon(0)\sim N(0, 4)$.

	\item[\bf (C2)] $\P(G=1\mid X) = \text{expit}\{(2X_1+ 2X_2 + 2X_1 X_2 - X_1^2)/4\}$, $\P(A=1\mid X, G) = \text{expit}\{(2X_1 - 2X_2 + 2X_1 X_2 - X_1^2  )/4 + G\}$, $Y(1) = 3 + 2 X_1 + X_2 + \epsilon(1)$, $Y(0) = 1 - 2 X_1 + 3 X_2 + \epsilon(0)$,  where $\epsilon(1)$ and $\epsilon(1)$ are generated the same as case (C1).

  \item[\bf (C3)] $\P(G=1\mid X) = \text{expit}\{(X_1+X_2)/2\}$, $\P(A=1\mid X, G) = \text{expit}\{(X_1 - X_2)/2 + G\}$, $Y(1) = 3 + 2 X_1 + X_2 + X_1^2 + \epsilon(1)$, $Y(0) = 1 - 2 X_1 + 3 X_2 + X_1 X_2 + \epsilon(0)$,  where  where $\epsilon(1)$ and $\epsilon(1)$ are generated the same as case (C1). 	
\end{itemize}

Assumptions \ref{assump1}-\ref{assump3} holds for all cases (C1)-(C3) in Study I. Thus, we can present the estimates $\hat \tau_{\mathrm{I}}$  and $\hat \tau_{\mathrm{VI}}$, but not $\hat \tau_{\mathrm{V}}$, as setting V is incompatible with Assumption \ref{assump3}. 
In addition, these three cases help to assess the doubly robustness of the proposed estimators as they can be categorized as follows: (C1) all working models are correctly specified; (C2) the outcome regression model is correctly specified, but the sampling score and the propensity score models are misspecified; (C3) the outcome regression model is misspecified, but the sampling score and the propensity score models are correctly specified.

Cases (C1)-(C3) compare the estimators $\hat \tau_{\mathrm{I}}$ and $\hat \tau_{\mathrm{VI}}$.  
To further compare them with $\hat \tau_{\mathrm{V}}$,  we set three additional simulation cases (C4)-(C6). Cases (C4)-(C6) have the same data-generation mechanism as cases (C1)-(C3) except that setting $\P(A=1\mid X, G = 0) = 0$.  Since cases (C4)-(C6) do not alter either the data-generating mechanism of the source data or the covariates distribution of the target data, the estimator $\hat \tau_{\mathrm{I}}$ under cases (C4)-(C6) remains the same as those under cases (C1)-(C3), respectively. This ensures a relatively fair comparison among $\hat \tau_{\mathrm{I}}$, $\hat \tau_{\mathrm{V}}$, and $\hat \tau_{\mathrm{VI}}$.

Each simulation study is based on 1000 replicates. In the following tables,
Bias and SD are the Monte Carlo bias and standard deviation over the
1000 simulations of the points estimates. CP95 is coverage proportions of the 95\%
confidence intervals based on the plug-in method described below Theorem \ref{thm3-3} in Section \ref{sec5}. 

\begin{table}[ht]
\caption{Study  I, efficiency comparison results for $\hat \tau_{\mathrm{I}}$, $\hat \tau_{\mathrm{V}}$, and $\hat \tau_{\mathrm{VI}}$.}
\centering
 \begin{tabular}{ccrrrrrrrrr}
  \toprule
   & & \multicolumn{3}{c}{$n =500$}    &  \multicolumn{3}{c}{$n =1000$}  &  \multicolumn{3}{c}{$n =2000$}    \\
  Case & Estimate & Bias & SD & CP95 &  Bias & SD & CP95  &  Bias & SD & CP95  \\ 
  \hline
(C1) &  $\hat \tau_{\mathrm{I}}$ &   0.007 & 0.502 & 0.942 & 0.008 & 0.347 & 0.942 & 0.001 & 0.246 & 0.947 \\ 
  (C1) & $\hat \tau_{\mathrm{VI}}$   & 0.013 & 0.356 & 0.957 & 0.004 & 0.245 & 0.946 & -0.002 & 0.180 & 0.941 \\ 
    (C4) & $\hat \tau_{\mathrm{V}}$   & 0.012 & 0.369 & 0.945 & 0.004 & 0.252 & 0.952 & 0.001 & 0.184 & 0.957  \\ \hdashline
  (C2) & $\hat \tau_{\mathrm{I}}$   & 0.009 & 0.498 & 0.941 & 0.018 & 0.352 & 0.947 & -0.005 & 0.241 & 0.954 \\ 
  (C2) &  $\hat \tau_{\mathrm{VI}}$   & -0.002 & 0.368 & 0.938 & 0.012 & 0.255 & 0.950 & -0.004 & 0.181 & 0.949 \\ 
    (C5) &  $\hat \tau_{\mathrm{V}}$   & -0.005 & 0.392 & 0.949 & 0.011 & 0.266 & 0.956 & 0.000 & 0.187 & 0.956  \\  \hdashline
  (C3) & $\hat \tau_{\mathrm{I}}$   & -0.009 & 0.634 & 0.957 & -0.004 & 0.435 & 0.962 & 0.009 & 0.307 & 0.960 \\ 
  (C3) &  $\hat \tau_{\mathrm{VI}}$   & -0.000 & 0.369 & 0.974 & -0.003 & 0.259 & 0.969 & -0.001 & 0.190 & 0.966 \\ 
    (C6) &  $\hat \tau_{\mathrm{V}}$   &  -0.006 & 0.441 & 0.968 & -0.012 & 0.299 & 0.975 & -0.001 & 0.228 & 0.976 \\ 
   \bottomrule
\end{tabular}  \label{tab2}
\end{table}

Table \ref{tab2} summarizes the numerical results of the proposed estimators $\hat \tau_{\mathrm{I}}$ and $\hat \tau_{\mathrm{VI}}$ for cases (C1)-(C3), as well as $\hat \tau_{\mathrm{V}}$ for cases (C4)-(C6).  
From Table \ref{tab2}, we have the following observations: (1) The Bias is small for all cases, demonstrating the double robustness of the proposed estimators;  (2) As expected, $\hat \tau_{\mathrm{V}}$ and $\hat \tau_{\mathrm{VI}}$ have significantly smaller SD than $\hat \tau_{\mathrm{I}}$, verifying the efficiency improvement by collecting extra data on $(A, Y)$ in the target data as shown in Theorems \ref{thm2}(a) and \ref{thm2}(b); 
    (3) The CP95 is close to its nominal value of 0.95 for cases (C1)-(C2) and (C4)-(C5) but only  slightly exceeds 0.95 for cases (C3) and (C6), indicating the stability of asymptotic variance estimation using the plug-in method;   (4) $\hat \tau_{\mathrm{V}}$ and $\hat \tau_{\mathrm{VI}}$ have the similar performance in terms of both Bias and SD. 
    This happens to be this result; as shown in Theorem \ref{thm2}(c), there is no deterministic order relation for their asymptotic variances and we will demonstrate this numerically in  Studies II and III.

\bigskip \noindent 
{\bf Study II.} 
We explore the efficiency gain and loss of $\tau$ under various sampling score models.  
    Two extra data-generating scenarios are given as follows: 
\begin{itemize}
	\item[\bf (C7)] 
 All data are generated as in (C1), expect 
 $\P(G=1\mid X) = \text{expit}\{(X_1+X_2 + 3)/2\}$.
 
    \item[\bf (C8)] All data are generated as in (C1), expect $\P(G=1\mid X) = \text{expit}\{(X_1+X_2 - 3)/2\}$. 
 \end{itemize}

  \begin{figure}[ht]
 \centerline{
 \includegraphics[scale = 1]{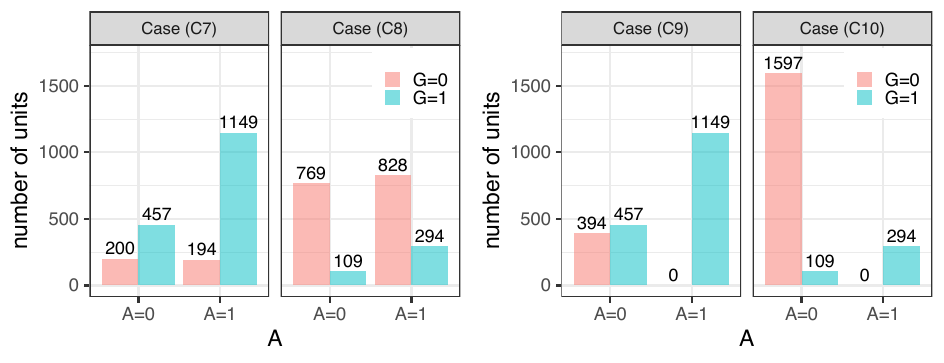}} 
 \vspace{-0.6cm}
 \caption{Bar plots for cases (C7)-(C10), based on one simulation ($n = 2000$).} 
 \label{fig1}
\end{figure}  

Similar to Study I, we also set simulation scenarios (C9)-(C10), which share the same data-generation mechanism as cases (C7)-(C8) with the only difference being that $\P(A=1\mid X, G = 0) = 0$. This allows us to compare the estimators $\hat \tau_{\mathrm{VI}}$ and $\hat \tau_{\mathrm{V}}$.

Cases (C7)-(C10) exhibit significant differences in their sampling score models, resulting in variations in the ratios, including $\gamma_1(X)$, $\alpha_2(X)$, $\alpha_3(X)$, $\gamma_2(X)$, and $\gamma_3(X)$ defined in Theorem \ref{thm2}.  Figure \ref{fig1} displays marginal distribution of $(A, G)$,  offering an intuitive visualization of the distinctions of the marginal ratios among cases (C7)-(C10).


\begin{table}[ht] 
\caption{Study II,  simulation studies under various sampling score models.}
\centering
 \begin{tabular}{ccrrrrrrrrr}
  \toprule
   & & \multicolumn{3}{c}{$n =500$}    &  \multicolumn{3}{c}{$n =1000$}  &  \multicolumn{3}{c}{$n =2000$}    \\
  Case & Estimate & Bias & SD & CP95 &  Bias & SD & CP95  &  Bias & SD & CP95  \\ 
  \hline
  (C7) & $\hat \tau_{\mathrm{I}}$  & -0.007 & 0.546 & 0.952 & 0.004 & 0.374 & 0.964 & -0.006 & 0.258 & 0.967 \\ 
  (C7) & $\hat \tau_{\mathrm{VI}}$ & 0.010 & 0.493 & 0.954 & 0.009 & 0.350 & 0.958 & -0.006 & 0.238 & 0.958 \\ 
  (C9) & $\hat \tau_{\mathrm{V}}$ & 0.009 & 0.486 & 0.955 & 0.004 & 0.344 & 0.957 & -0.008 & 0.235 & 0.966 \\ \hdashline 
  (C8) & $\hat \tau_{\mathrm{I}}$ & 0.033 & 0.702 & 0.898 & 0.003 & 0.480 & 0.924 & 0.000 & 0.321 & 0.950 \\ 
  (C8) & $\hat \tau_{\mathrm{VI}}$ & 0.013 & 0.303 & 0.948 & 0.007 & 0.208 & 0.957 & -0.005 & 0.146 & 0.953 \\
   (C10) & $\hat\tau_{\mathrm{V}}$ & 0.012 & 0.407 & 0.947 & 0.007 & 0.283 & 0.935 & -0.003 & 0.195 & 0.941    \\
  \bottomrule
\end{tabular}  \label{tab3}
\end{table}

Table \ref{tab3} presents the simulation results for cases (C7)-(C10). In cases (C7) and (C9), $\hat \tau_{\mathrm{V}}$ exhibits the smallest SD,  slightly smaller than that of $\hat \tau_{\mathrm{VI}}$. The estimators $\hat \tau_{\mathrm{V}}$ and $\hat \tau_{\mathrm{VI}}$ only show a small improvement in SD compared to $\hat \tau_{\mathrm{I}}$, indicating a limited efficiency gain. This is mainly due to the truth that $\pi(X)$ tends to be large, while $\alpha(X)$, $\gamma_1(X)$, and $\gamma_2(X)$ tend to be small, as depicted in Figure \ref{fig1}. Consequently, as stated in Theorems \ref{thm2}(b) and \ref{thm2}(c), 
    the contribution to efficiency improvement is small in these cases.
In contrast, for cases (C8) and (C10), $\hat \tau_{\mathrm{VI}}$ demonstrates the smallest SD, and both $\hat \tau_{\mathrm{V}}$ and $\hat \tau_{\mathrm{VI}}$ exhibit significant improvements in SD compared to $\hat \tau_{\mathrm{I}}$. This suggests a substantial increase in efficiency. The primary reason for this lies in the fact that $\pi(X)$ tends to be small, while $\alpha(X)$, $\gamma_1(X)$, and $\gamma_2(X)$ tend to be large, as shown in Figure \ref{fig1} again.

 \bigskip \noindent 
{\bf Study III}.  
 According to Theorem \ref{thm2}, the potential outcome's variance also plays an important role in the efficiency gain. To explore this, we further conduct sensitivity analysis on the potential outcome's variance. Two extra data-generating scenarios are considered. 
\begin{itemize} 
\item[~~\bf (C11)]  All data are generated as in case (C1), except $\epsilon(1) \sim N(0, 1)$, $\epsilon(0) \sim N(0, 9)$.   

    \item[~~\bf (C12)] All data are generated as in case (C1), except $\epsilon(1) \sim N(0, 9)$, $\epsilon(0) \sim N(0, 1)$.  
 \end{itemize}

We introduce scenarios (C13)-(C14), mirroring cases (C11)-(C12) in data generation, except for $\P(A=1\mid X, G = 0) = 0$, to facilitate comparison with $\hat \tau_{\mathrm{V}}$.

\begin{table}[h!] 
\caption{Study III, sensitivity analysis on the potential outcome's variance.}
\centering
\begin{tabular}{ccrrrrrrrrr}
  \toprule
   & & \multicolumn{3}{c}{$n =500$}    &  \multicolumn{3}{c}{$n =1000$}  &  \multicolumn{3}{c}{$n =2000$}    \\
  Case & Estimate & Bias & SD & CP95 &  Bias & SD & CP95  &  Bias & SD & CP95  \\ \hline 
  (C11) & $\hat \tau_{\mathrm{I}}$ & 0.011 & 0.617 & 0.930 & 0.011 & 0.428 & 0.937 & 0.002 & 0.302 & 0.941 \\ 
  (C11) & $\hat \tau_{\mathrm{VI}}$ & 0.020 & 0.381 & 0.955 & 0.005 & 0.263 & 0.957 & 0.000 & 0.195 & 0.944 \\ 
  (C13) & $\hat \tau_{\mathrm{V}}$ & 0.018 & 0.350 & 0.951 & 0.006 & 0.241 & 0.957 & 0.001 & 0.179 & 0.947 \\ \hdashline 
  (C12) & $\hat \tau_{\mathrm{I}}$ & 0.004 & 0.452 & 0.949 & 0.005 & 0.309 & 0.952 & 0.001 & 0.220 & 0.959 \\ 
  (C12) & $\hat \tau_{\mathrm{VI}}$ & 0.007 & 0.357 & 0.953 & 0.003 & 0.247 & 0.953 & -0.004 & 0.179 & 0.953 \\ 
  (C14) & $\hat \tau_{\mathrm{V}}$ & 0.006 & 0.417 & 0.947 & 0.003 & 0.283 & 0.953 & 0.001 & 0.205 & 0.953 \\ 
   \bottomrule
\end{tabular} \label{tab4}
\end{table}

In Table \ref{tab4}, both $\hat \tau_{\mathrm{V}}$ and $\hat \tau_{\mathrm{VI}}$ exhibit efficiency gains over $\hat \tau_{\mathrm{I}}$. However, when comparing $\hat \tau_{\mathrm{V}}$ and $\hat \tau_{\mathrm{VI}}$,  we observe that $\hat \tau_{\mathrm{V}}$ performs relatively better when the variance of $\epsilon(0)$ is larger than that of $\epsilon(1)$. Conversely, it performs relatively worse when the variance of $\epsilon(0)$ is smaller than that of $\epsilon(1)$. This further demonstrates the conclusions in Theorem \ref{thm2}.

\bigskip \noindent 
{\bf Study IV}. We also consider more complex data generating mechanism and estimation via modern nonparametric regression methods. We consider three additional simulation cases below:  
\begin{itemize} 
    \item[\bf (C15)]   $\P(G=1| X) = \text{expit}\{(2X_1+2X_2 + 2X_1 X_2 -X_1^2)/4\}$, $\P(A=1| X, G) = \text{expit}\{(2X_1 - 2X_2+ 2X_1 X_2 -X_1^2)/4 + G\}$, $Y(1) = 3 + 2 X_1 + \exp(X_2) + (1-G)X_1/2 + \epsilon(1)$, $Y(0) = 1 - 2 X_1 + 3 X_2 + \log(|X_1|) + \epsilon(0)$,  where $\text{expit}(x) = \exp(x)/\{1 + \exp(x)\}$,  $X = (X_1, X_2)^{\intercal} \sim N(0, I_2)$, $\epsilon(1)\sim N(0, 4)$ and $\epsilon(0)\sim N(0, 4)$. 

    \item[\bf (C16)] 
$\P(G=1| X) = \text{expit}\{(2X_1+2X_2 + 2X_1 X_2 -X_1^2)/4\}$, $\P(A=1| X, G) = \text{expit}\{(2X_1 - 2X_2+ 2X_1 X_2 -X_1^2)/4 + G\}$, $Y(1) = 3 + 2 X_1 + sin(X_2) + X_1^2 + \epsilon(1)$, $Y(0) = 1 - 2 X_1 + 3 X_2 + X_1X_2 + \epsilon(0)$,  where $\text{expit}(x) = \exp(x)/\{1 + \exp(x)\}$,  $X = (X_1, X_2)^{\intercal} \sim N(0, I_2)$, $\epsilon(1)\sim N(0, 4)$ and $\epsilon(0)\sim N(0, 4)$. 

    \item[\bf (C17)]
$\P(G=1| X) = \text{expit}\{(2X_1+2X_2 + 2X_1 X_2 -X_1^2)/4\}$, $\P(A=1| X, G) = \text{expit}\{(2X_1 - 2X_2+ 2X_1 X_2 -X_1^2)/4 + G\}$, $Y(1) = 3 + 2 X_1 + X_2 + X_1^2 + \epsilon(1)$, $Y(0) = 1 - 2 X_1 + 3 X_2 + X_1X_2 + \epsilon(0)$,  where $\text{expit}(x) = \exp(x)/\{1 + \exp(x)\}$,  $X = (X_1, X_2)^{\intercal} \sim N(0, I_2)$, $\epsilon(1)\sim N(0, 4)$ and $\epsilon(0)\sim N(0, 4)$. 
\end{itemize}
Cases (C15)-(C17) have complex function forms of propensity score, sampling score, and outcome regression models.  
Similar to Study I, we also set simulation scenarios (C18)-(C20), which share the same data-generation mechanism as cases (C15)-(C17) with the only difference being that $\P(A=1\mid X, G = 0) = 0$. This allows us to compare the estimators $\hat \tau_{\mathrm{VI}}$ and $\hat \tau_{\mathrm{V}}$. The outcome variable \(Y\) is income in 1978 (in thousands of dollars).

We use random forest (R package \texttt{grf}) to estimate all the nuisance parameter with four-fold cross-fitting.  
Each simulation study is based on 1000 replicates. In the following tables, Bias and SD are the Monte Carlo bias and standard deviation over the
1000 simulations of the points estimates. CP95 is coverage proportions of the 95\% confidence intervals based on the plug-in method described below Theorem \ref{thm3} in Section \ref{sec5}.

\begin{table}[ht]
\caption{Study  IV, Efficiency comparison results for $\hat \tau_{\mathrm{I}}$, $\hat \tau_{\mathrm{V}}$, and $\hat \tau_{\mathrm{VI}}$.}
\centering
\begin{tabular}{ccrrrrrrrrr}
  \toprule
   & & \multicolumn{3}{c}{$n =500$}    &  \multicolumn{3}{c}{$n =1000$}  &  \multicolumn{3}{c}{$n =2000$}    \\
  Case & Estimate & Bias & SD & CP95 &  Bias & SD & CP95  &  Bias & SD & CP95  \\ 
  \hline
(C15) &  $\hat \tau_{\mathrm{I}}$ &0.021 & 0.702 & 0.965 & 0.064 & 0.487 & 0.965 & 0.075 & 0.330 & 0.953 \\ 
  (C15) & $\hat \tau_{\mathrm{VI}}$   &  -0.017 & 0.385 & 0.962 & 0.001 & 0.276 & 0.960 & 0.019 & 0.196 & 0.961 \\ 
    (C18) & $\hat \tau_{\mathrm{V}}$   & 0.129 & 0.405 & 0.944 & 0.090 & 0.287 & 0.942 & 0.083 & 0.208 & 0.933  \\ \hdashline
  (C16) & $\hat \tau_{\mathrm{I}}$   & -0.206 & 0.780 & 0.920 & -0.115 & 0.540 & 0.933 & -0.086 & 0.360 & 0.927 \\ 
  (C16) &  $\hat \tau_{\mathrm{VI}}$   &-0.133 & 0.394 & 0.948 & -0.090 & 0.278 & 0.948 & -0.074 & 0.200 & 0.942 \\ 
    (C19) &  $\hat \tau_{\mathrm{V}}$   & -0.136 & 0.407 & 0.930 & -0.115 & 0.282 & 0.931 & -0.090 & 0.208 & 0.930  \\  \hdashline
  (C17) & $\hat \tau_{\mathrm{I}}$   & -0.206 & 0.779 & 0.918 & -0.112 & 0.539 & 0.936 & -0.083 & 0.359 & 0.921 \\ 
  (C17) &  $\hat \tau_{\mathrm{VI}}$   & -0.143 & 0.388 & 0.949 & -0.095 & 0.273 & 0.948 & -0.078 & 0.196 & 0.942 \\ 
    (C20) &  $\hat \tau_{\mathrm{V}}$   & -0.132 & 0.403 & 0.932 & -0.109 & 0.279 & 0.936 & -0.084 & 0.206 & 0.935 \\ 
   \bottomrule
\end{tabular} \label{tab-res1}
\end{table}

Table \ref{tab-res1} summarizes the numerical results of the proposed estimators $\hat \tau_{\mathrm{I}}$ and $\hat \tau_{\mathrm{VI}}$ for cases (C15)-(C17), as well as $\hat \tau_{\mathrm{V}}$ for cases (C18)-(C20). From Table \ref{tab-res1}, one can see that $\hat \tau_{\mathrm{V}}$ and $\hat \tau_{\mathrm{VI}}$ still have significantly smaller SD than $\hat \tau_{\mathrm{I}}$, verifying the efficiency improvement by collecting extra data on $(A, Y)$ in the target data.

\section{Application}
\label{sec:job-app} 
   We demonstrate the application of our proposed methodology through two examples. The first example in Section \ref{ssec:job-app}  uses source dataset from a randomized experiment with the structure \((A,X,Y)\) and target dataset from an observational study with the structure \((A=0,X,Y)\). The second example in Section \ref{2sec:app}  involves both source and target datasets from observational studies, with the data structure \((A,X,Y)\) \citep{Chu-Yang2023}.

    \subsection{Application to job training program}
    \label{ssec:job-app}
 The first analysis uses experimental data from the National Supported Work (NSW) program and non-experimental data from the Population Survey of Income Dynamics (PSID) \citep{lalonde1986evaluating,Wu2023Transfer}. The PSID data serves as the target sample, and the NSW data serves as the source sample. The NSW dataset includes 185 individuals who received job training (\(A = 1\)) and 260 individuals who did not (\(A = 0\)), while the PSID dataset consists of 128 individuals who did not receive job training. The eight collected baseline variables \( X \) include age, education, race, Hispanic status, marital status, high school degree, and incomes for 1974 and 1975.   The outcome variable \(Y\) is income in 1978 (in thousands of dollars). 
 
 Table \ref{tab: summary-stat-job} presents the summary statistics for the baseline covariates and outcome, revealing differences between the two datasets.   We find that the individuals in the NSW data are relatively younger compared to those in the PSID data. The proportion of married individuals is lower in the NSW data, at 0.19 for the treated group and 0.15 for the control group, compared to a higher proportion of 0.70 in the PSID data. Income levels for individuals in the NSW data are relatively lower for 1974 and 1975 compared to those in the PSID data. The sample means for the outcome variable $Y$ are 6.35 and 4.55 for the treated and control groups in the NSW data, respectively, while the mean in the PSID data is 5.28. 

\begin{table}[t]
    \caption{Mean and standard deviations (in parenthesis) of baseline characteristics.}
    \label{tab: summary-stat-job}
    \centering\resizebox{0.8074\textwidth}{!}{
\begin{tabular}{ccccccc}
\toprule
       Variables   &  & \begin{tabular}[c]{@{}c@{}} Treated  Group  \\ in Source Data\end{tabular} &  & \begin{tabular}[c]{@{}c@{}}Control Group  \\ in Source Data \end{tabular} &  & \begin{tabular}[c]{@{}c@{}}Control Group  \\ in Target Data\end{tabular} \\   \cline{1-1} \cline{3-3} \cline{5-5} \cline{7-7} \addlinespace[2mm]
Age  ($X_1$)     &  & 25.82 (7.16)                                                              &  & 25.05 (7.06)                                                                 &  & 38.26 (12.89)                                                                \\
Education  ($X_2$)   &  & 10.35 (2.01)                                                              &  & 10.09 (1.61)                                                                 &  & 10.30 (3.18)                                                                  \\
Race   ($X_3$)    &  & 0.84 (0.36)                                                               &  & 0.83 (0.38)                                                                  &  & 0.45 (0.50)                                                                   \\
Hispanic   ($X_4$)  &  & 0.06 (0.24)                                                               &  & 0.11 (0.31)                                                                  &  & 0.12 (0.32)                                                                  \\
Marital Status   ($X_5$)  &  & 0.19 (0.39)                                                               &  & 0.15 (0.36)                                                                  &  & 0.70 (0.46)                                                                   \\
Nodegree   ($X_6$)  &  & 0.71 (0.46)                                                               &  & 0.83 (0.37)                                                                  &  & 0.51 (0.5)                                                                   \\
Income74     ($X_7$)   &  & 2.10 (4.89)                                                                &  & 2.11 (5.69)                                                                  &  & 5.57 (7.26)                                                                  \\
Income75    ($X_8$)    &  & 1.53 (3.22)                                                               &  & 1.27 (3.10)                                                                   &  & 2.61 (5.57)                                                                  \\
Income78 ($Y$)       &  & 6.35 (7.87)                                                               &  & 4.55 (5.48)                                                                  &  & 5.28 (7.76)                                                                  \\ \bottomrule
\end{tabular}} 
\end{table}

We are interested in whether the job training program can increase income for the target population in 1978. Since the source data is an experimental dataset, Assumption \ref{assump1} naturally holds. We perform the analysis under Assumption \ref{assump2} and consider two estimation methods based on the EIFs in Theorem \ref{thm1}: the first method uses linear models for the outcome  models and logistic regression for the propensity scores and sampling scores, with estimators denoted as \(\hat{\tau}_{\mathrm{I}}\) and \(\hat{\tau}_{\mathrm{V}}\), as detailed in Section \ref{sec5}. The second method employs nonparametric estimation techniques, using random forest for the outcome  models and neural networks for the propensity scores and sampling scores, with estimators denoted as \(\hat{\tau}_{\mathrm{I}}^{\mathrm{np}}\) and \(\hat{\tau}_{\mathrm{V}}^{\mathrm{np}}\). 
 Table \ref{res:job-trianing} presents the estimation  results  for the target population, including point estimates, variance estimates, and 95\% confidence intervals (CIs). The point estimates are very close, suggesting that the job training program could increase income in the target dataset by approximately 2.5 to 3. Additionally, the estimation variance in Scenario IV is smaller than in Scenario V for both methods, further validating Theorem \ref{thm2}(b). All 95\% CIs, except for the estimator $\hat\tau_{\mathrm{V}}$ under Setting V, include zero. However, since the lower bounds of the 95\% CIs are   close to zero, these empirical results almost support the conclusion that the job training program would increase income in the target dataset.

\begin{table}
\caption{Point estimate, variance, and 95\% confidence interval under different scenarios.} 
\label{res:job-trianing}
\centering
\resizebox{0.940\textwidth}{!}{
\begin{tabular}{ccccccccc}
\toprule
      Methods                             &                   & Settings                         &  & Point Estimate &  & Variance &  & 95\% Confidence Interval \\ \cline{1-1} \cline{3-9} 
\multirow{2}{*}{Parametric Models} & \multirow{2}{*}{} & I-IV ($\hat \tau_{\mathrm{I}}$)  &  & 2.98           &  & 2.37     &  & (-0.04, 6.00)            \\
                                   &                   & V ($\hat \tau_{\mathrm{V}}$)     &  & 3.54           &  & 2.28     &  & (0.60, 6.50)            \\ \cline{1-1} \cline{3-9} 
 \multirow{2}{*}{Nonparametric Models} & \multirow{2}{*}{} & I-IV ($\hat \tau_{\mathrm{I}}^{\mathrm{np}}$)  &  & 2.52            &  &  2.25    &  &    (-0.42, 5.46)        \\
                                   &                   & V ($\hat \tau_{\mathrm{V}}^{\mathrm{np}}$)     &  &  2.49           &  &  2.23     &  &  (-0.43, 5.42)         \\ 
 \bottomrule
\end{tabular}}
\end{table}
\begin{figure}[t]
    \centering
    \includegraphics[width=0.95\linewidth]{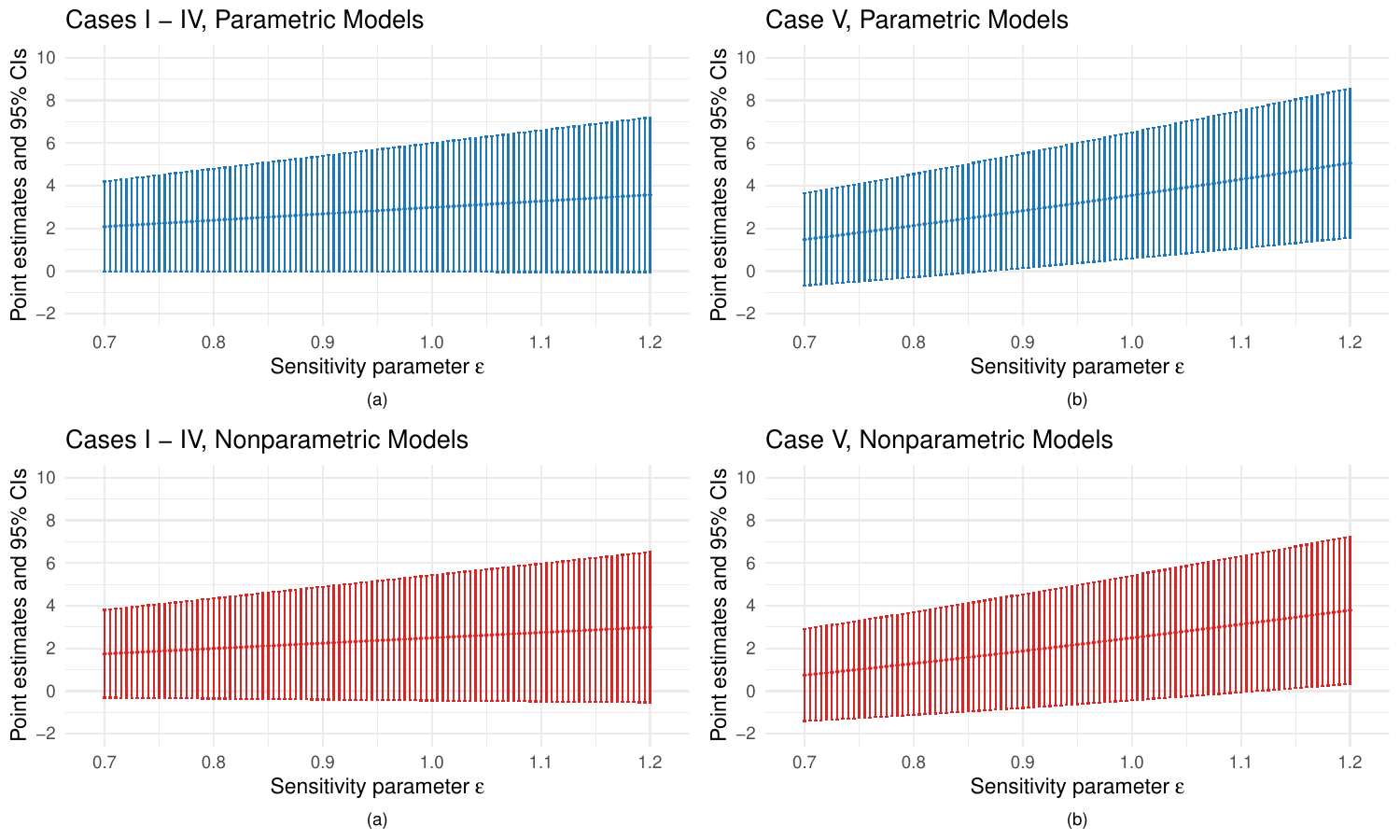}
    \caption{Sensitivity analysis for various cases.}
    \label{fig:square}
\end{figure}

{Next, we address potential violations of transportability between the two datasets using the method proposed in Section \ref{sec:sen}, by applying a series of linear drift functions \(\psi_0(u) = \epsilon u\) for some known $\epsilon$, as specified in Assumption \ref{assump5}, for sensitivity analysis. 
  Under the linear drift function setting \(\psi_0(u)=\epsilon u\), we have: 
\begin{equation}
    \label{eq:post0}
    \epsilon := \frac{\tilde{\mu}_0(X)}{\mu_0(X)} = \frac{\E\{Y(0) \mid X, G=0\}}{\E\{Y(0) \mid X, G=1\}} .
\end{equation}
To estimate the empirical range of \(\epsilon\), we regress the fitted values of \(\E(Y \mid A=0, G=0, X)\) on \(\E(Y \mid A=0, G=1, X)\), resulting in an empirical bound for \(\epsilon\) in \eqref{eq:post0} of \([0.7, 1.2]\).  We then conduct sensitivity analysis within this range. Specifically, for each given \(\epsilon\) in this range, we estimate \(\tau\) using the estimators proposed in Section \ref{est-postdrift}. We first consider parametric models for all nuisance functions, using linear models for outcome regression and logistic regression for both propensity and sampling scores. We  set the ratios \(r_0^\circ(X) = r_1^\circ(X) = 1\), which results in \(r_0(X) = r_1(X) = \epsilon^2\). While correctly specifying \(r_0(X)\) and \(r_1(X)\) may be challenging, this does not impact the estimation consistency as shown in Section \ref{est-postdrift}.  The estimation results are presented in the first row of Figure \ref{fig:square}. We observe that as \(\epsilon\) increases, point estimates under settings I and V also increase. Additionally, as \(\epsilon\) increases, the 95\% CIs for settings I-IV nearly all exclude zero, and the  95\% CIs for setting V remain significantly above zero.  
We also perform sensitivity analysis using nonparametric techniques, where all nuisance models are estimated using random forests and neural networks.} The results, shown in the second row of Figure \ref{fig:square}, are consistent with the trends observed in the parametric models, and we omit the discussion here. In summary, the sensitivity analysis results in Figure \ref{fig:square} indicate that point estimates and 95\% CIs for \(\tau\) are relatively robust to the choice of sensitivity parameters $\epsilon$. These findings suggest that the job training program is likely to improve incomes in the target population  \citep{lalonde1986evaluating,dehejia2002propensity,Wu2023Transfer}.
\subsection{Application to the MIMIC-III data}
\label{2sec:app}
We also demonstrate the application of our proposed methodology using data derived from two distinct clinical databases: the MIMIC-III clinical database and the eICU collaborative research database. In this context, the MIMIC-III data serves as our target sample, while the eICU data acts as our source sample.  Both datasets focus on patients with sepsis,  but they collected from different hospitals and time periods, indicating the potential presence of heterogeneity.  For the MIMIC-III dataset, we screened the patients based on the Sepsis-3 criteria, which was developed by the Society of Critical Care Medicine and the European Society of Intensive Care Medicine \citep{singer2016third}, and is currently the primary criteria for sepsis diagnosis. The specific extraction code used for extracting data from the MIMIC-III dataset was the official visualization view code provided by \href{https://github.com/MIT-LCP/mimic-code}{MIT-LCP/mimic-code}. Similar selection methods can be found in previous studies \citep{verboom2019robustness,zhu2024effect,chen2021association,popoff2021effect}.  For the eICU dataset, we similarly employed the Sepsis-3 criteria for patient screening.  Therefore, the proposed method is applied in this context to assess the causal effects in the target population by integrating these two samples.


\begin{table}[h]
    \caption{Descriptions, means, and standard deviations (in parenthesis) of baseline characteristics in the source and target samples.}  
    \label{tab: summary-stat}
    \centering\resizebox{0.9995043465\textwidth}{!}{
\begin{tabular}{|c|c|c|c|}
\hline
{ Variables}                            & { Descriptions}                                                         & { Source Data}                     & { Target Data}                     \\ \hline
\multirow{2}{*}{Vasopressor $A$}      & The patient received one of the treatments among norepinephrine,    & \multirow{2}{*}{0.26 (0.44)}    & \multirow{2}{*}{0.42 (0.49)}    \\
                                      & epinephrine, dopamine, dobutamine, and vasopressin.                  &                                 &                                 \\ \hline
\multirow{2}{*}{$\log\vert$Cumulative Balance$\vert$ $Y$ }      & Cumulative Balance is calculated by subtracting the fluid output 48 hours  & \multirow{2}{*}{   7.07 (1.22)   }    & \multirow{2}{*}{7.19 (1.09)   }    \\
                                      & after admission from the fluid input at the time of initial admission.                 &                                 &                                 \\ \hline
Age (years) $X_1$                     & The patient's age in years.                                          & 65.52 (14.63) & 64.5 (17.62)                \\ \hline
Gender $X_2$                          & The patient's gender.                                                & 0.52 (0.5) & 0.57 (0.5)                \\ \hline
Weight  (Kg) $X_3$                    & The patient's weight.                                                &  87.27 (30.56) & 82.89 (24.71)                 \\ \hline
\multirow{2}{*}{Temperature (°C) $X_4$         }     & The patient's temperature in degrees Celsius,                        & \multirow{2}{*}{36.90 (0.70)}   & \multirow{2}{*}{36.88 (0.68)}   \\
                                      & averaged over the first day in the ICU.                              &                                 &                                 \\ \hline
\multirow{2}{*}{Glucose (mg/dL) $X_5$ }      & The patient's glucose level in milligrams per deciliter,             & \multirow{2}{*}{151.67 (51.56)} & \multirow{2}{*}{145.69 (49.09)} \\
                                      & averaged over the first day in the ICU.                              &                                 &                                 \\ \hline
\multirow{2}{*}{Creatinine (mg/dL)  $X_6$}   & The patient's creatinine level in milligrams per deciliter,          & \multirow{2}{*}{ 1.75 (1.61)}    & \multirow{2}{*}{1.47 (1.44)}    \\
                                      & averaged over the first day in the ICU.                              &                                 &                                 \\ \hline
\multirow{2}{*}{BUN (mg/dL) $X_7$}          & The patient's blood urea nitrogen level in milligrams per deciliter, & \multirow{2}{*}{32.83 (22.74)}  & \multirow{2}{*}{28.11 (21.83)}  \\
                                      & averaged over the first day in the ICU.                              &                                 &                                 \\ \hline
\multirow{2}{*}{WBC (K/uL) $X_8$}           & The patient's white blood cell count in thousands per microliter,    & \multirow{2}{*}{14.12 (8.98)}   & \multirow{2}{*}{12.91 (7.41)}   \\
                                      & averaged over the first day in the ICU.                              &                                 &                                 \\ \hline
\multirow{2}{*}{PCO2 (mm Hg) $X_9$}         & The patient's partial pressure of carbon dioxide in millimeters      & \multirow{2}{*}{44.15 (15.37)}  & \multirow{2}{*}{41.1 (10.08)}  \\
                                      & of mercury,  averaged over the first day in the ICU.                 &                                 &                                 \\ \hline
\multirow{2}{*}{pH (units) $X_{10}$}           & The patient's pH level,                                              & \multirow{2}{*}{7.35 (0.09)}    & \multirow{2}{*}{7.36 (0.08)}    \\
                                      & averaged over the first day in the ICU.                              &                                 &                                 \\ \hline
\multirow{2}{*}{Chloride (mmol/L) $X_{11}$}       & The patient's chloride level in millimoles per liter,                & \multirow{2}{*}{103.81 (7.02)}  & \multirow{2}{*}{ 104.73 (6.21)}  \\
                                      & averaged over the first day in the ICU.                              &                                 &                                 \\ \hline
\multirow{2}{*}{Potassium (mmol/L) $X_{12}$}      & The patient's potassium level in millimoles per liter,               & \multirow{2}{*}{4.18 (0.63)}    & \multirow{2}{*}{4.20 (0.65)}    \\
                                      & averaged over the first day in the ICU.                              &                                 &                                 \\ \hline
\multirow{2}{*}{Sodium (mmol/L)  $X_{13}$}         & The patient's sodium level in millimoles per liter,                  & \multirow{2}{*}{138.34 (5.49)}  & \multirow{2}{*}{138.24 (4.82) }  \\
                                      & averaged over the first day in the ICU.                              &                                 &                                 \\ \hline
\multirow{2}{*}{SpO2 (\%) $X_{14}$}               & The patient's oxygen saturation level as a percentage,               & \multirow{2}{*}{96.55 (2.63)}   & \multirow{2}{*}{ 97.01 (2.15) }   \\
                                      & averaged over the first day in the ICU.                              &                                 &                                 \\ \hline
\multirow{2}{*}{Heart Rate (bpm) $X_{15}$}        & The patient's heart rate in beats per minute,                        & \multirow{2}{*}{ 91.47 (17.26)}  & \multirow{2}{*}{87.94 (16.12)}  \\
                                      & averaged over the first day in the ICU.                              &                                 &                                 \\ \hline
\end{tabular}
}
    \label{tab:cov_dep}
\end{table}

We denote patients who received one of the following vasopressor treatments as $A=1$: norepinephrine, epinephrine, dopamine, dobutamine, or vasopressin. Conversely, $A=0$ corresponds to patients who did not receive any of these vasopressor treatments. Similar treatment variable definitions can be found in previous studies \citep{zhu2024effect, avni2015vasopressors, buchtele2022applicability, shi2020vasopressors, jentzer2015pharmacotherapy}. Our primary focus is on the cumulative fluid balance (measured in milliliters), calculated by subtracting the fluid output 48 hours after admission from the fluid input at the time of initial admission. A positive cumulative balance indicates that a patient's fluid input exceeds their output, a condition known as hypervolemia or fluid overload. In critically ill patients, fluid overload is associated with an increased mortality rate and various complications, including pulmonary edema, heart failure, tissue damage, and impaired bowel function \citep{claure2016fluid}. Conversely, a negative cumulative balance suggests that a patient's fluid output exceeds their input, a condition known as hypovolemia. Severe hypovolemia can lead to ischemic injury in vital organs, resulting in multi-system organ failure \citep{taghavi2023hypovolemic}. Therefore, we denote the outcome as $Y= \log\vert$Cumulative Balance$\vert$, where lower values of the outcome variable are considered preferable, and a similar way of outcome variable can be seen in \citet{Chu-Yang2023}. 
We consider 15 baseline covariates denoted as $X=(X_1,\ldots,X_{15})$ in both datasets. These covariates include age ($X_1$), gender ($X_2$), weight  ($X_3$), temperature  ($X_4$),  glucose levels ($X_5$), creatinine levels ($X_6$), blood urea nitrogen levels ($X_7$), white blood cell counts ($X_8$), partial pressure of carbon dioxide ($X_9$), pH levels ($X_{10}$), chloride levels ($X_{11}$), potassium levels ($X_{12}$), sodium levels ($X_{13}$), oxygen saturation levels ($X_{14}$), and heart rates ($X_{15}$). These covariates offer a comprehensive overview of various aspects of the patient's health and physiological status.  Table \ref{tab:cov_dep} provides specific descriptions, sample means, and sample variances of these variables.  Minor differences can be observed in certain covariates, such as glucose levels $X_5$, blood urea nitrogen levels $X_7$, and white blood cell counts $X_8$. Moreover, the outcome variable $Y$ also exhibits a minor difference. After removing missing and abnormal values, the MIMIC-III dataset comprises 1763 subjects, while the eICU dataset comprises 3403 individuals.

\begin{table}[t]
\caption{Point estimate, standard error, and 95\% confidence interval under different scenarios.} 
\label{res:mimic3}
\centering
\resizebox{0.906818\textwidth}{!}{
\begin{tabular}{ccccccccc}
\toprule\addlinespace[0.5mm]
                                   &                   & Settings                         &  & Point Estimate &  & Standard Error &  & 95\% Confidence Interval \\ \addlinespace[0.5mm]\cline{1-1} \cline{3-9} \addlinespace[0.5mm]
\multirow{2}{*}{Parametric Models} & \multirow{2}{*}{} & I-IV ($\hat \tau_{\mathrm{I}}$)  &  & 0.04           &  &0.06    &  & (-0.07,  0.15 )            \\
                                   &                   & V ($\hat \tau_{\mathrm{V}}$)     &  & 0.05       &  & 0.04     &  & (-0.03, 0.12)            \\ \addlinespace[0.5mm]\cline{1-1} \cline{3-9} \addlinespace[0.5mm]
 \multirow{2}{*}{Nonparametric Models} & \multirow{2}{*}{} & I-IV ($\hat \tau_{\mathrm{I}}^{\mathrm{np}}$)  &  & 0.04           &  &  0.05    &  &    (-0.06, 0.15)        \\
                                   &                   & V ($\hat \tau_{\mathrm{V}}^{\mathrm{np}}$)     &  &   0.04     &  &  0.04   &  &  (-0.03, 0.12)         \\ \addlinespace[0.5mm]
 \bottomrule
\end{tabular}}
\end{table}

While we have complete access to the target sample, our ability to utilize all the collected information can be constrained by different assumptions. Next, we will evaluate the causal effects on the target dataset under various scenarios, as outlined in Table \ref{tab:6settings} of the main text. As demonstrated in Theorem \ref{thm1}, under certain assumptions, Cases I-IV (when the target dataset lacks ignorability) will result in the exact same EIF and estimation process. We consider two approaches for estimating the working models: the first uses parametric models, and the second employs nonparametric models. 

In the first case of parametric models, we use two linear models to estimate the conditional outcome means \(\mu_0(X)\) and \(\mu_1(X)\) for Cases I-IV. For all cases, we use logistic regression models to estimate the propensity scores \(e_0(X)\) and \(e_1(X)\), as well as the sampling score \(\pi(X)\). In Case VI, we similarly use linear models to estimate the conditional outcome means \(\tilde\mu_0(X)\) and \(\tilde\mu_1(X)\).

In the second case of nonparametric models, we use random forests to estimate the conditional outcome means \(\mu_0(X)\) and \(\mu_1(X)\) for Cases I-IV. For all cases, we use neural networks to model the propensity scores \(e_0(X)\) and \(e_1(X)\), as well as the sampling score \(\pi(X)\). In Case VI, we also use random forests to estimate the conditional outcome means \(\tilde\mu_0(X)\) and \(\tilde\mu_1(X)\).

Table \ref{res:mimic3} presents the point estimates and corresponding 95\% confidence intervals for the ATE under Assumption \ref{assump2}. The estimates from both parametric and nonparametric models are similar. Under the parametric model, the point estimate for Cases I-IV is 0.04, but the 95\% confidence interval includes 0, making the result insignificant. Similarly, the point estimate obtained under setting VI suggests that the use of vasopressors does not significantly affect cumulative balance. Moreover, we observe a smaller variance in Case VI compared to Cases I-IV, which is consistent with the insights provided by Theorem \ref{thm1}.
 
\begin{figure}[t]
    \centering
    \includegraphics[width=0.905\linewidth]{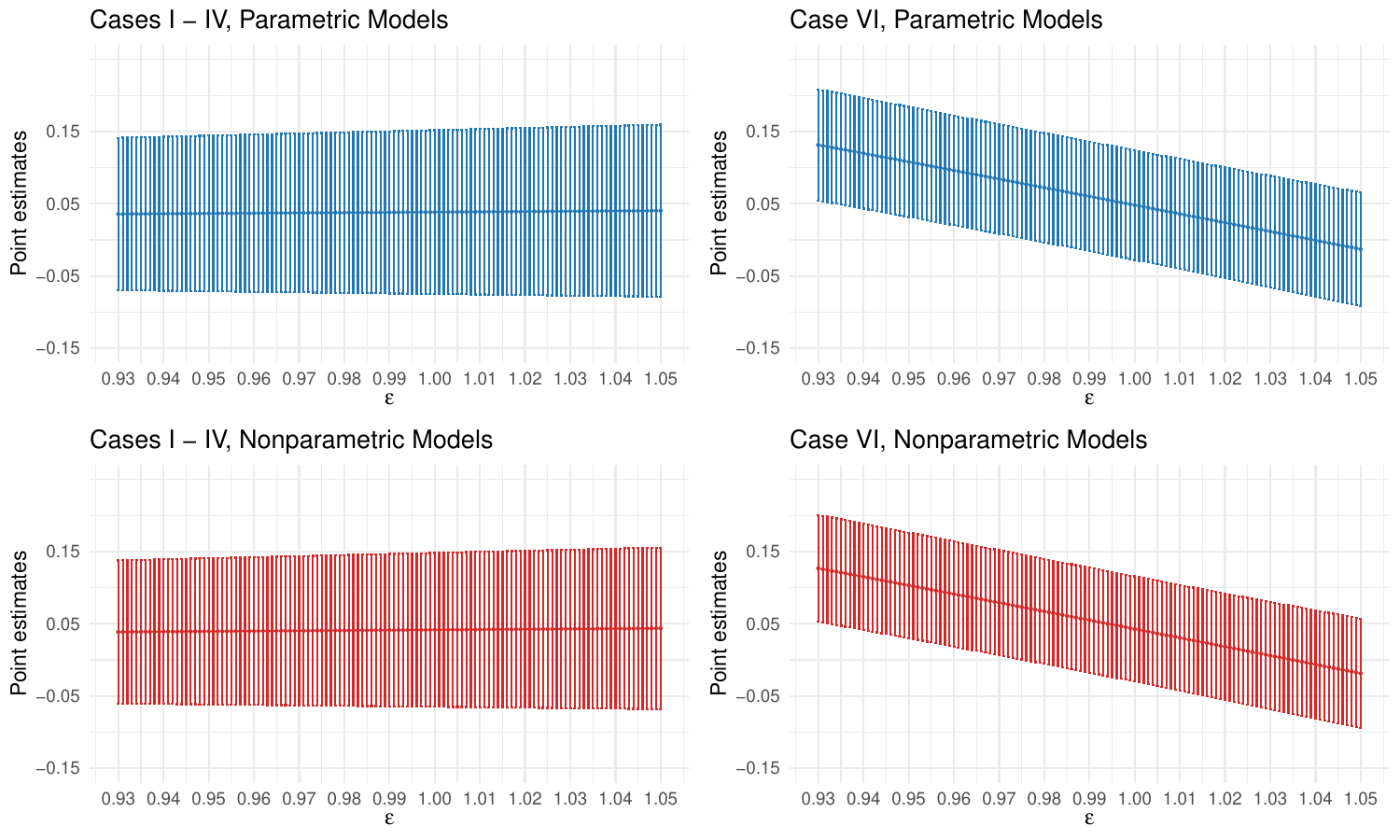}
    \caption{Sensitivity analysis for various cases.}
    \label{fig:liner-mimic3} 
\end{figure} 
Next, we address the posterior drift problem using the method outlined in Section \ref{sec:sen} of the main text. First, within the class of linear drift functions, we focus on the scenario where:
$$\psi_{0}(u)= \epsilon u ,~ \psi_{1}(u)=\epsilon u .$$
Therefore, for these linear drift functions, equation \eqref{eq:sensitivity-pars} holds. The empirical range of \(\epsilon\) can be calculated as \([0.93, 1.05]\), and we will use this range to conduct sensitivity analyses separately for Cases I-IV and Case VI. In this sensitivity analysis, we apply both parametric and nonparametric models to the working models, consistent with the settings in the previous subsection. The estimated results under various sensitivity parameters are presented in Figure \ref{fig:liner-mimic3}. The results from the parametric models are very similar to those from the nonparametric models, and we interpret the results based on the parametric models.

We observe that as \(\epsilon\) increases, the length of the 95\% confidence intervals also gradually increases. Additionally, all \ intervals for Cases I-IV contain zero, indicating that the causal effect for the target population is not significant. Similarly, almost all confidence intervals for Case VI contain zero, further validating the nonsignificant treatment effect of vasopressors in the target population. Finally, we find that the length of the 95\% confidence intervals for Case VI is significantly smaller than for Cases I-IV, which aligns well with the conclusion provided by Theorem \ref{thm5}.

In summary, the results in Figure \ref{fig:liner-mimic3} highlight the importance of conducting sensitivity analysis in observational studies, as introducing Assumption \ref{assump3} may lead to significant changes in the results. However, we still observe that most of the 95\% confidence intervals for the estimates include zero, indicating that vasopressors do not have a significant effect on the outcome variable. Additionally, as vasopressors are generally an effective clinical treatment, the likelihood of observing significant side effects is low \citep{zhu2024effect, avni2015vasopressors, buchtele2022applicability, shi2020vasopressors, jentzer2015pharmacotherapy}.


%
\section{Extension to ATE in the Source Population}   
 \label{sec:extension-ATE-source}
In the previous sections, we focus on the efficiency comparative analysis on $\tau$ across various settings.  Now, we turn our attention to a similar analysis of the estimand $\beta = \E[ Y(1) - Y(0) \mid G= 1]$, which also may be of great interest in many real-world application scenarios. 

 \subsection{Efficiency Comparison}
 
 Different from $\tau$, $\beta$ is identifiable based solely on the source data in all settings $\mathrm{I}$-$\mathrm{VI}$, incorporating additional target data is only to enhance the estimation efficiency. In parallel to Theorem \ref{thm1}, we obtain the EIFs and efficiency bounds for $\beta$, as shown in Theorem \ref{thm6}.

\begin{theorem}[efficiency bound of $\beta$]  \label{thm6}   The following statements hold: 

(a) the efficient influence function of $\beta$ under settings $\mathrm{I}$, $\mathrm{II}$, $\mathrm{III}$, and $\mathrm{IV}$ is given as
	\begin{align*}
	\varphi_{\mathrm{I}} =   \varphi_{\mathrm{II}} =  \varphi_{\mathrm{III}} ={}& \varphi_{\mathrm{IV}} =  \frac{G}{q} \left[  \frac{A\{Y - \mu_1(X)\}}{e_1(X)} - \frac{(1-A)\{Y - \mu_0(X)\}}{1 - e_1(X)} + \{\mu_1(X) - \mu_0(X) - \beta\} \right], 
	\end{align*}
 which also is the efficient influence function based only the source data. 
The associated semiparametric efficiency bound is $\mathbb{\bar V}_{\mathrm{I}}^* = \mathbb{\bar  V}_{\mathrm{II}}^*  =\mathbb{\bar  V}_{\mathrm{III}}^* =\mathbb{\bar V}_{\mathrm{IV}}^*  =  \E(\varphi_{\mathrm{I}}^2)$.

(b)  the efficient influence function of $\beta$ under setting $\mathrm{V}$  is given as
	\begin{align*}
	\varphi_\mathrm{V} ={}&  \frac{G}{q} \left[  \frac{A\{Y - \mu_1(X)\}}{e_1(X)}  + \{\mu_1(X) - \mu_0(X) - \beta\} \right]  - \frac{1}{q} \frac{(1-A)\{Y - \mu_0(X)\} \pi(X) }{\tilde \P(A=0\mid X)},
	\end{align*}
The efficiency bound is $\mathbb{\bar  V}_{\mathrm{V}}^* =  \E (\varphi_{\mathrm{V}}^2)$.

(c) the efficient influence function of $\beta$ under setting $\mathrm{VI}$ is given as
			\begin{align*}
	\varphi_{\mathrm{VI}} 
	={}& \frac{1}{q}  \left ( \frac{A\{Y - \mu_1(X)\}  \pi(X) }{ \P(A=1\mid X)}  -  \frac{(1-A)\{Y - \mu_0(X)\} \pi(X) }{ \P(A=0 \mid X)}  \right )  +  \frac{G}{q}\{\mu_1(X) -  \mu_0(X) -  \beta \}, 
	\end{align*} 
The semiparametric efficiency bound is $\mathbb{\bar  V}_{\mathrm{VI}}^* =  \E (\varphi_{\mathrm{VI}}^2)$.		
	In particular, 
when $e_0(X) = 0$, $\varphi_{\mathrm{VI}}$ reduces to $\varphi_{\mathrm{V}}$;    
when $\pi(X) = 1$, $\varphi_{\mathrm{VI}}$ reduces to $\varphi_{\mathrm{I}}$, the efficient influence function based only the source data. 
\end{theorem}

Under Assumptions \ref{assump1}-\ref{assump2}, Theorem \ref{thm3}(a) shows that the EIF of $\beta$ remains unchanged with the gathering of additional target data. This implies that the extra data contributes no information to the inference on $\beta$. 
Similarly, in line with Theorems \ref{thm2}(b) and \ref{thm2}(c), Theorems \ref{thm3}(b) and \ref{thm3}(c) reveal that the additional target data can be leveraged to enhance the efficiency of $\beta$ when Assumption \ref{assump3} or \ref{assump4} is further imposed.

\begin{theorem}[efficiency comparison of $\beta$] \label{thm7} We have that 

(a) the semiparametric efficiency bounds of $\beta$ under settings $\mathrm{I}$-$\mathrm{VI}$ satisfy that 
    \begin{align*}  
     \mathbb{\bar V}_{\mathrm{I}}^* = \mathbb{\bar V}_{\mathrm{II}}^*  =\mathbb{\bar V}_{\mathrm{III}}^* =\mathbb{\bar V}_{\mathrm{IV}}^*  >  \mathbb{\bar V}_{\mathrm{V}}^*, \quad \mathbb{\bar V}_{\mathrm{I}}^* = \mathbb{\bar V}_{\mathrm{II}}^*  =\mathbb{\bar V}_{\mathrm{III}}^* =\mathbb{\bar V}_{\mathrm{IV}}^*  > \mathbb{\bar V}_{\mathrm{VI}}^*. 
    \end{align*}

  (b) the efficiency gain of setting $\mathrm{V}$ compared to setting I is 
    \begin{align*}
    \mathbb{\bar V}_\mathrm{I}^* - \mathbb{\bar V}_\mathrm{V}^* =  \frac{1}{q^2} \E \left [ \gamma(X) \pi(X) \frac{ \mathrm{Var}\{Y(0) \mid X\}}{ 1 - e_1(X) } \right ],
\end{align*}
where $\gamma(X)$ is defined in Theorem \ref{thm2}(b). 


(c) the efficiency gain of setting $\mathrm{VI}$ compared to setting I is
\begin{align*}
    \mathbb{\bar V}_\mathrm{I}^* - \mathbb{\bar V}_{\mathrm{VI}}^* =  \frac{1}{q^2} \E \left [ \frac{ \alpha_1(X) \pi(X) \mathrm{Var}\{Y(1) \mid X\}}{ e_1(X) } +  \frac{ \gamma_1(X) \pi(X)\mathrm{Var}\{Y(0) \mid X\}}{ 1 - e_1(X) } \right ],
\end{align*}
where $\alpha_1(X)$ and $\gamma_1(X)$ are defined in Theorem \ref{thm2}(c).

(d) the difference between $\mathbb{\bar V}_\mathrm{V}^*$ and $\mathbb{\bar V}_{\mathrm{VI}}^*$ is given as 
 \begin{align*}
    \mathbb{\bar V}_\mathrm{V}^* - \mathbb{\bar V}_{\mathrm{VI}}^*
    ={}&  \frac{1}{q^2} \E \left [ \frac{\alpha_2(X) \pi^2(X)  \mathrm{Var}\{Y(1) \mid X\}}{ \tilde \P(A=1\mid X) } - \frac{\gamma_2(X) \pi^2(X) \mathrm{Var}\{Y(0) \mid X\} }{\tilde \P(A=0\mid X)}  \right ],
\end{align*}
where  $\alpha_2(X)$ and $\gamma_2(X)$ are defined in Theorem \ref{thm2}(d).
	 		 
\end{theorem}


 
 Theorem \ref{thm4} provides a comparison of efficiency bounds for $\beta$ across settings $\mathrm{I}$-$\mathrm{VI}$, with most conclusions aligning with those in Theorem \ref{thm2}, except for the observation that the efficiency gain of $\beta$ increases as $\pi(X)$ increases, which is contrary to the efficiency gain of $\tau$. 
Moreover, when Assumption \ref{assump2} is replaced by Assumption \ref{assump5}, we further present the corresponding EIFs and efficiency bounds of $\beta$. 

\begin{theorem}[efficiency bound of $\beta$ when Assumption \ref{assump2} is replaced by Assumption \ref{assump5}]  \label{thm6-v2}   The following statements hold: 

(a) the efficient influence function of $\beta$ under settings $\mathrm{I}^*$, $\mathrm{II}^*$, $\mathrm{III}^*$, and $\mathrm{IV}^*$ is given as 
	\begin{align*}
	\varphi_{\mathrm{I}^*} =   \varphi_{\mathrm{II}^*} =  \varphi_{\mathrm{III}^*} ={}& \varphi_{\mathrm{IV}^*} =  \frac{G}{q} \left[  \frac{A\{Y - \mu_1(X)\}}{e_1(X)} - \frac{(1-A)\{Y - \mu_0(X)\}}{1 - e_1(X)} + \{\mu_1(X) - \mu_0(X) - \beta\} \right], 
	\end{align*}
 which also is the efficient influence function based only the source data. 
The associated semiparametric efficiency bound is $\mathbb{\bar V}_{\mathrm{I}^*}^* = \mathbb{\bar  V}_{\mathrm{II}^*}^*  =\mathbb{\bar  V}_{\mathrm{III}^*}^* =\mathbb{\bar V}_{\mathrm{IV}^*}^*  =  \E(\varphi_{\mathrm{I}^*}^2)$.

(b)  the efficient influence function of $\beta$ under setting $\mathrm{V}^*$  is given as
	\begin{align*}
	\varphi_{\mathrm{V}^*} ={}& \frac{G\pi(X)}{q}  \left ( \frac{A\{Y - \mu_1(X)\}    }{\tilde \P_w(A=1\mid X)}  -  \frac{(1-A)\{Y - \mu_0(X)\}  }{\tilde \P_w(A=0 \mid X)}  \right ) \\
	{}& -  \frac{(1-G)\pi(X)}{q}    \frac{(1-A)\{Y - \tilde \mu_0(X)\} r_0(X)  }{ \psi_0'(\mu_0(X)) \tilde \P_w(A=0 \mid X)}    +  \frac{G}{q}\{\mu_1(X) -  \mu_0(X) -  \beta \}, 
	\end{align*}
 where $\tilde \P_w(A=1|X) = \pi(X)e_1(X)$ and $\tilde \P_w(A=0|X) = \pi(X)\{1-e_1(X)\} +  \{1-\pi(X)\} r_0(X)$. 
The efficiency bound is $\mathbb{\bar  V}_{\mathrm{V}^*}^* =  \E (\varphi_{\mathrm{V}^*}^2)$.

(c) the efficient influence function of $\beta$ under setting $\mathrm{VI}^*$ is given as
			\begin{align*}
	\varphi_{\mathrm{VI}^*} 
	={}& \frac{G\pi(X)}{q}  \left ( \frac{A\{Y - \mu_1(X)\}    }{ \P_w(A=1\mid X)}  -  \frac{(1-A)\{Y - \mu_0(X)\}  }{ \P_w(A=0 \mid X)}  \right ) \\
	{}& +   \frac{(1-G)\pi(X)}{q}  \left ( \frac{A\{Y - \tilde \mu_1(X)\}  r_1(X)  }{\psi_1'(\mu_1(X))  \P_w(A=1\mid X)}  -  \frac{(1-A)\{Y -\tilde \mu_0(X)\} r_0(X)  }{ \psi_0'(\mu_0(X))  \P_w(A=0 \mid X)}  \right )  \\
	{}&   +  \frac{G}{q}\{\mu_1(X) -  \mu_0(X) -  \beta \}, 
	\end{align*} 
where $\P_w(A=1|X) = \pi(X) e_1(X) + (1- \pi(X)) e_0(X) r_1(X)$ and $\P_w(A=0|X) = \pi(X)(1 - e_1(X)) + (1-\pi(X))(1 - e_0(X)) r_0(X)$. 	
The semiparametric efficiency bound is $\mathbb{\bar  V}_{\mathrm{VI}^*}^* =  \E (\varphi_{\mathrm{VI}^*}^2)$.		
	In particular, 
when $e_0(X) = 0$, $\varphi_{\mathrm{VI}^*}$ reduces to $\varphi_{\mathrm{V}^*}$;    
when $\pi(X) = 1$, $\varphi_{\mathrm{VI}^*}$ reduces to $\varphi_{\mathrm{I}^*}$, the efficient influence function based only the source data. 
\end{theorem}

Under Assumptions \ref{assump1} and \ref{assump5}, Theorem \ref{thm6-v2}(a) shows that the EIF of $\beta$ remains unchanged with the gathering of additional target data. This implies that the extra data contributes no information to the inference on $\beta$.
 Similarly, in line with Theorems \ref{thm5}, Theorems \ref{thm6-v2}(b) and \ref{thm6-v2}(c) reveal that the additional target data can be leveraged to enhance the efficiency of $\beta$ when Assumption \ref{assump3} or \ref{assump4} is further imposed. 
 We then compare the efficiency bounds in Theorem \ref{thm6-v2}.

\begin{theorem}[efficiency comparison of $\beta$  when Assumption \ref{assump2} is replaced by Assumption \ref{assump5}] \label{thm7-v2} We have that 
 the semiparametric efficiency bounds of $\beta$ under settings $\mathrm{I}^*$-$\mathrm{VI}^*$ satisfy that 
    \begin{align*}  
     \mathbb{\bar V}_{\mathrm{I}^*}^* = \mathbb{\bar V}_{\mathrm{II}^*}^*  =\mathbb{\bar V}_{\mathrm{III}^*}^* =\mathbb{\bar V}_{\mathrm{IV}^*}^*  >  \mathbb{\bar V}_{\mathrm{V}^*}^*, \quad \mathbb{\bar V}_{\mathrm{I}^*}^* = \mathbb{\bar V}_{\mathrm{II}^*}^*  =\mathbb{\bar V}_{\mathrm{III}^*}^* =\mathbb{\bar V}_{\mathrm{IV}^*}^*  > \mathbb{\bar V}_{\mathrm{VI}^*}^*. 
    \end{align*}

  (a) the efficiency gain of setting $\mathrm{V}^*$ compared to setting $\mathrm{I}^*$ is 
    \begin{align*}
    \mathbb{\bar V}_{\mathrm{I}^*}^* - \mathbb{\bar V}_{\mathrm{V}^*}^* =  \frac{1}{q^2} \E \left [\tilde \gamma_1(X) \pi(X) \frac{ \mathrm{Var}\{Y(0) \mid X, G=1\}}{ 1 - e_1(X) } \right ],
\end{align*}
where $\tilde \gamma_1(X)$ is defined in Theorem \ref{thm5}(a). 


(b) the efficiency gain of setting $\mathrm{VI}^*$ compared to setting $\mathrm{I}^*$ is
\begin{align*}
    \mathbb{\bar V}_{\mathrm{I}^*}^* - \mathbb{\bar V}_{\mathrm{VI}^*}^* =  \frac{1}{q^2} \E \left [ \frac{ \tilde \alpha_2(X) \pi(X) \mathrm{Var}\{Y(1) \mid X, G=1\}}{ e_1(X) } +  \frac{ \tilde \gamma_2(X) \pi(X)\mathrm{Var}\{Y(0) \mid X, G=1\}}{ 1 - e_1(X) } \right ],
\end{align*}
where $\tilde \alpha_2(X)$ and $\tilde \gamma_2(X)$ are defined in Theorem \ref{thm5}(b).

(c) the difference between $\mathbb{\bar V}_{\mathrm{V}^*}^*$ and $\mathbb{\bar V}_{\mathrm{VI}^*}^*$ is given as 
 \begin{align*}
    \mathbb{\bar V}_{\mathrm{V}^*}^* - \mathbb{\bar V}_{\mathrm{VI}^*}^*
    ={}&  \frac{1}{q^2} \E \left [ \frac{\tilde \alpha_3(X) \pi^2(X)  \mathrm{Var}\{Y(1) \mid X, G=1\}}{ \tilde \P_w(A=1\mid X) } - \frac{\tilde \gamma_3(X) \pi^2(X) \mathrm{Var}\{Y(0) \mid X, G=1\} }{\tilde \P_w(A=0\mid X)}  \right ],
\end{align*}
where  $\tilde \alpha_3(X)$ and $\tilde \gamma_3(X)$ are defined in Theorem \ref{thm5}(c), and  $\tilde \P_w(A=1|X) = \pi(X)e_1(X)$ and $\tilde \P_w(A=0|X) = \pi(X)\{1-e_1(X)\} +  \{1-\pi(X)\} r_0(X)$. 
	 		 
\end{theorem}

\subsection{Estimation and Inference}

Similar to the estimation of $\tau$ in the manuscript, the proposed estimators of $\beta$ under settings I and VI are   
\begin{align*}  
\hat \beta_{\mathrm{I}} ={}& \P_n \left [ \frac{G}{q} \left \{  \frac{A\{Y - \hat \mu_1(X)\}}{\hat e_1(X)} - \frac{(1-A)\{Y - \hat \mu_0(X)\}}{1 - \hat e_1(X)} + \{ \hat  \mu_1(X) - \hat  \mu_0(X)\} \right \} \right ],  \\
\hat \beta_{\mathrm{VI}} ={}& \P_n  \left [ \frac{1}{q}  \left ( \frac{A(Y - \hat  \mu_1(X)) \hat   \pi(X) }{\hat  e(X) }  -  \frac{(1-A)(Y -\hat  \mu_0(X)) \hat  \pi(X) }{1-\hat  e(X) } \right )  +  \frac{G}{q} \{ \hat  \mu_1(X) - \hat  \mu_0(X)\} \right ],   
\end{align*} 
and for setting $\mathrm{V}$, $\hat \beta_\mathrm{V}$ is of the same form as $\hat \beta_{\mathrm{VI}}$, except that $\hat e_0(X)$ is replaced with 0. Also, we establish the asymptotic properties for the estimators of $\beta$ in Theorem \ref{thm12}. 

   \begin{theorem} 
\label{thm12}
Suppose that the regularity conditions considered in   \cite{newey1994large} hold, then 
 
(a) under Assumptions \ref{assump1}-\ref{assump2}, the estimator $\hat \beta_{\mathrm{I}}$ is consistent and asymptotically normal if either (i) $\mu_a(X)$ for $a = 0, 1$ are correctly specified, or (ii) $e_1(X)$ is correctly specified.  



(b) under Assumptions \ref{assump1}-\ref{assump2}  and \ref{assump4},  $\hat \beta_\mathrm{V}$ is consistent and asymptotically normal if either (i) $\mu_a(X)$ for $a = 0, 1$ are correctly specified, or (ii) $e_1(X)$ and $\pi(X)$ are correctly specified.  


(c)  under Assumptions \ref{assump1}-\ref{assump3}, $\hat \beta_{\mathrm{VI}}$ is consistent and asymptotically normal if (i) $\mu_a(X)$ for $a = 0, 1$ are correctly specified, or (ii)  $e_1(X)$, $e_0(X)$, and $\pi(X)$ are correctly specified. 


(d) all the proposed estimators of $\tau$ achieve the corresponding semiparametric efficiency bounds when all the working models are correctly specified. 

\end{theorem}

\subsection{Simulation Studies} \label{simu-beta}
We also conduct numerical experiments to assess the finite-sample performance of the proposed estimators for  $\beta$ in all cases (C1)-(C14). The corresponding results are similar to those of the estimators for $\tau$ in the main text.

\begin{table}[h!] 
\caption{Study I, efficiency comparison between $\hat \beta_\mathrm{I}$, $\hat \beta_\mathrm{V}$, and $\hat \beta_\mathrm{VI}$.}
\centering
\begin{tabular}{ccrrrrrrrrr}
  \toprule
   & & \multicolumn{3}{c}{$n =500$}    &  \multicolumn{3}{c}{$n =1000$}  &  \multicolumn{3}{c}{$n =2000$}    \\
  Case & Estimate & Bias & SD & CP95 &  Bias & SD & CP95  &  Bias & SD & CP95  \\ 
  \hline
(C1) & $\hat \beta_\mathrm{I}$ & 0.003 & 0.441 & 0.938 & 0.013 & 0.301 & 0.959 & 0.002 & 0.207 & 0.959 \\ 
  (C1) & $\hat \beta_\mathrm{VI}$ & 0.005 & 0.360 & 0.951 & 0.008 & 0.252 & 0.957 & 0.001 & 0.172 & 0.965 \\ 
  (C4) &  $\hat \beta_\mathrm{V}$ & 0.009 & 0.353 & 0.950 & 0.008 & 0.246 & 0.959 & 0.003 & 0.168 & 0.958 \\   \hdashline
  (C2) & $\hat \beta_\mathrm{I}$ & 0.012 & 0.405 & 0.959 & 0.003 & 0.286 & 0.964 & -0.006 & 0.195 & 0.967 \\ 
  (C2) & $\hat \beta_\mathrm{VI}$ & -0.003 & 0.320 & 0.969 & -0.002 & 0.232 & 0.958 & -0.003 & 0.163 & 0.963 \\ 
  (C5) &  $\hat \beta_\mathrm{V}$ & 0.002 & 0.312 & 0.971 & 0.001 & 0.226 & 0.966 & 0.000 & 0.165 & 0.956 \\  \hdashline
  (C3) & $\hat \beta_\mathrm{I}$ & -0.012 & 0.507 & 0.935 & 0.004 & 0.343 & 0.957 & 0.008 & 0.240 & 0.965 \\ 
  (C3) & $\hat \beta_\mathrm{VI}$ & 0.001 & 0.398 & 0.957 & 0.005 & 0.279 & 0.962 & 0.005 & 0.192 & 0.969  \\ 
  (C6) & $\hat \beta_\mathrm{V}$ & 0.007 & 0.396 & 0.947 & 0.009 & 0.273 & 0.960 & 0.006 & 0.189 & 0.955 \\ 
   \bottomrule
\end{tabular} \label{tab-s1}
\end{table}


\begin{table}[h!] 
\caption{Study II, sensitivity analysis on sampling score.}
\centering
\begin{tabular}{ccrrrrrrrrr}
  \toprule
   & & \multicolumn{3}{c}{$n =500$}    &  \multicolumn{3}{c}{$n =1000$}  &  \multicolumn{3}{c}{$n =2000$}    \\
  Case & Estimate & Bias & SD & CP95 &  Bias & SD & CP95  &  Bias & SD & CP95  \\ 
  \hline
(C7) & $\hat \beta_\mathrm{I}$ & -0.005 & 0.338 & 0.952 & 0.009 & 0.236 & 0.952 & 0.001 & 0.170 & 0.937 \\ 
  (C7) & $\hat \beta_\mathrm{VI}$ & 0.005 & 0.306 & 0.951 & 0.011 & 0.220 & 0.945 & 0.001 & 0.158 & 0.943 \\ 
  (C9) &  $\hat \beta_\mathrm{V}$ & 0.007 & 0.301 & 0.946 & 0.008 & 0.215 & 0.943 & -0.001 & 0.152 & 0.937 \\ \hdashline
  (C8) & $\hat \beta_\mathrm{I}$ & 0.007 & 0.655 & 0.957 & -0.004 & 0.470 & 0.962 & -0.004 & 0.330 & 0.967 \\ 
  (C8) & $\hat \beta_\mathrm{VI}$ & -0.009 & 0.471 & 0.974 & -0.004 & 0.337 & 0.980 & -0.007 & 0.249 & 0.972 \\ 
  (C10) & $\hat \beta_\mathrm{V}$ & -0.012 & 0.496 & 0.977 & -0.003 & 0.355 & 0.971 & -0.004 & 0.261 & 0.973 \\
   \bottomrule
\end{tabular} \label{tab-s2}
\end{table}

\begin{table}[h!] 
\caption{Study III, sensitivity analysis on the potential outcome's variance.}
\centering
\begin{tabular}{ccrrrrrrrrr}
  \toprule
   & & \multicolumn{3}{c}{$n =500$}    &  \multicolumn{3}{c}{$n =1000$}  &  \multicolumn{3}{c}{$n =2000$}    \\
  Case & Estimate & Bias & SD & CP95 &  Bias & SD & CP95  &  Bias & SD & CP95  \\   \hline 
  (C11) & $\hat \beta_\mathrm{I}$ & 0.001 & 0.527 & 0.936 & 0.016 & 0.360 & 0.958 & 0.001 & 0.248 & 0.960 \\ 
  (C11) & $\hat \beta_\mathrm{VI}$ & 0.005 & 0.393 & 0.954 & 0.008 & 0.278 & 0.951 & 0.002 & 0.191 & 0.967 \\ 
  (C13) & $\hat \beta_\mathrm{V}$ & 0.010 & 0.356 & 0.956 & 0.008 & 0.250 & 0.956 & 0.002 & 0.170 & 0.961 \\  \hdashline
  (C12) & $\hat \beta_\mathrm{I}$ & 0.006 & 0.402 & 0.941 & 0.010 & 0.276 & 0.955 & 0.002 & 0.190 & 0.960 \\ 
  (C12) & $\hat \beta_\mathrm{VI}$ & 0.005 & 0.355 & 0.954 & 0.008 & 0.248 & 0.958 & 0.000 & 0.170 & 0.961 \\ 
  (C14) & $\hat \beta_\mathrm{V}$ & 0.009 & 0.377 & 0.952 & 0.007 & 0.262 & 0.957 & 0.003 & 0.180 & 0.963 \\ 
   \bottomrule
\end{tabular} \label{tab-s3}
\end{table}

\section{Extension to ATT}  
\label{sec:ext-ATT}
In this section, we further extend the efficiency comparison for the average treatment effect on the treated (ATT) in both the target population and the source population, respectively.

\subsection{ATT in the Target Population}  \label{sec4-1-suppl}

Let $\tau_{att} = \E[Y(1) - Y(0) \mid G=0, A = 1]$,  it is easy to see that $\tau_{att}$ is not identifiable for settings I, III, and V. Thus, we discuss the efficiency gain and loss for $\tau_{att}$ under settings  $\mathrm{II}$, $\mathrm{IV}$, and $\mathrm{VI}$. Let $e_0 = \E[e_0(X)] =\P(A=1|G=0)$.

\begin{theorem}[efficiency bound of $\tau_{att}$]    \label{thm1-suppl}
The following statements hold: 

(a) the efficient influence function of $\tau_{att}$ under settings $\mathrm{II}$ and $\mathrm{IV}$  is given as
	\begin{align*}
	 \phi _{att, \mathrm{II}}  ={}& \phi_{att, \mathrm{IV}} =  \frac{G}{(1-q)e_0} \left \{  \frac{A(Y - \mu_1(X))}{e_1(X)} - \frac{(1-A)(Y - \mu_0(X))}{1 - e_1(X)} \right \} \frac{(1 - \pi(X))e_0(X)}{\pi(X)} \\ 
	+{}&  \frac{(1-G)A}{(1-q)e_0} ( \mu_1(X) - \mu_0(X) - \tau_{att}). 
	\end{align*}
The associated semiparametric efficiency bound is $\mathbb{V}_{att, \mathrm{II}}^* = \mathbb{V}_{att, \mathrm{IV}}^*  =  \E[\phi_{att, \mathrm{II}}^2]$.

(b)  the efficient influence function of $\tau$ under setting $\mathrm{VI}$ is given as
			\begin{align*}
	\phi_{att, \mathrm{VI}} ={}& \frac{(1 - \pi(X)) e_0(X) }{(1-q)e_0}  \left ( \frac{A(Y - \mu_1(X)) }{\P(A=1|X)}  -  \frac{(1-A)(Y - \mu_0(X)) }{ \P(A=0|X) } \right )  \\
		+{}&  \frac{(1-G)A}{(1-q)e_0} ( \mu_1(X) - \mu_0(X) - \tau_{att}). 
	\end{align*}
The semiparametric efficiency bound is $\mathbb{V}_{att, \mathrm{VI}}^* =  \E[\phi_{att, \mathrm{VI}}^2]$.		
	In particular,  when $e_1(X) = e_0(X) := e(X)$, $\phi_{att, \mathrm{VI}}$ has a simpler form given by
		\begin{align*}
 & \frac{1}{(1-q)e_0}  \left ( \frac{A(Y - \mu_1(X))  }{ e(X) }  -  \frac{(1-A)(Y - \mu_0(X))}{ 1 - e(X) }  \right ) (1 - \pi(X)) e(X) \\
 +{}& \frac{(1-G)A}{(1-q)e_0} ( \mu_1(X) - \mu_0(X) - \tau_{att}). 
	\end{align*} 

(c) When $\pi(X) = 0$, $\phi_{att, \mathrm{VI}}$ reduces to the efficient influence function based only the target data given by
    		\begin{align*}
	\phi_{att} ={}& \frac{1-G}{1-q} \left ( \frac{A(Y - \mu_1(X)) }{ e_0 }  -  \frac{(1-A)(Y - \mu_0(X)) e_0(X) }{ e_0 (1 - e_0(X))} + \frac{A}{e_0} ( \mu_1(X) - \mu_0(X) - \tau_{att}) \right ),
	\end{align*}
 its efficient influence bound is denoted as $\mathbb{V}_{att}^* =  \E[\phi_{att}^2]$.	
 
\end{theorem}

\begin{theorem}[efficiency comparison of $\tau_{att}$] \label{thm2-suppl} We have that 

(a) the semiparametric efficiency bounds of $\tau_{att}$ under settings  $\mathrm{II}$, $\mathrm{IV}$, and $\mathrm{VI}$ satisfy 
    \begin{align*}
        \mathbb{V}_{att, \mathrm{II}}^* = \mathbb{V}_{att, \mathrm{IV}}^* > \mathbb{V}_{att, \mathrm{VI}}^*.
    \end{align*}
In addition, the efficiency gain of setting $\mathrm{VI}$ compared to setting II is 
\begin{align*}
      \V_{att, \mathrm{II}}^* - \V_{att, \mathrm{VI}}^*  = \E \Biggl [ \frac{ (1-\pi(X))^2  e_0(X)^2   }{ \pi(X) (1-q)^2 e_0^2 }  \Big \{  \alpha_1(X)  \cdot \frac{\text{Var}( Y(1) | X)}{e_1(X) }  + \gamma_1(X) \cdot  \frac{ \text{Var}( Y(0) | X) }{(1 - e_1(X) )} \Big \} \Biggr  ],
\end{align*}
where $\alpha_1(X)$ and $\gamma_1(X)$ are defined in Theorem \ref{thm2}(b). 

(b)  the efficiency gain from leveraging the source data is 
    \[ \V_{att}^* - \V_{att, \mathrm{VI}}^*  =\E \left [  \frac{ (1 - \pi(X)) e_0(X)}{(1-q)^2 e_0^2}    \left \{ \alpha_3(X) \text{Var}( Y(1) | X) +   \gamma_3(X) \frac{\text{Var}( Y(0) | X) e_0(X)}{(1 - e_0(X) )} \right \}  \right ],  \]
   where $\alpha_3(X)$ and $\gamma_3(X)$ are defined in Proposition \ref{prop1}. 
\end{theorem}


\subsection{ATT in the Source Population}  \label{sec4-2-suppl}

Let $\beta_{att} = \E[Y(1) - Y(0) \mid G=1, A = 1]$ and $e_1 = \E[e_1(X)] =\P(A=1|G=1)$. 

\begin{theorem}[efficiency bound of $\beta_{att}$]  \label{thm3-suppl}  The following statements hold: 

(a) the efficient influence function of $\beta_{att}$ under settings $\mathrm{I}$, $\mathrm{II}$, $\mathrm{III}$, and $\mathrm{IV}$ is given as 
	\begin{align*}
  & 	\varphi_{att, \mathrm{I}} =   \varphi_{att, \mathrm{II}} =  \varphi_{att, \mathrm{III}} = \varphi_{att, \mathrm{IV}}  \\
 ={}&  \frac{G}{q e_1} \left \{  \frac{A(Y - \mu_1(X))}{e_1(X)} - \frac{(1-A)(Y - \mu_0(X))}{1 - e_1(X)} \right \} e_1(X) + \frac{G A}{q e_1} \left ( \mu_1(X) - \mu_0(X) - \beta_{att} \right ), 
	\end{align*}
 which also is the efficient influence function based only the source data. 
The associated semiparametric efficiency bound is $\mathbb{\bar V}_{att, \mathrm{I}}^* = \mathbb{\bar  V}_{att, \mathrm{II}}^*  =\mathbb{\bar  V}_{att, \mathrm{III}}^* =\mathbb{\bar V}_{att, \mathrm{IV}}^*  =  \E[\varphi_{att, \mathrm{I}}^2]$.

(b)  the efficient influence function of $\beta_{att}$ under setting $\mathrm{V}$  is given as
	\begin{align*}
	\varphi_{att, \mathrm{V}} ={}&  \frac{G}{qe_1} A(Y - \mu_1(X))  - \frac{1}{qe_1} \frac{(1-A)(Y - \mu_0(X)) \pi(X) e_1(X) }{ \tilde \P(A=0|X) } + \frac{G A}{q e_1} \left ( \mu_1(X) - \mu_0(X) - \beta_{att} \right ).
	\end{align*}
The semiparametric efficiency bound is $\mathbb{\bar  V}_{att, \mathrm{V}}^* =  \E[\varphi_{att, \mathrm{V}}^2]$.

(c) the efficient influence function of $\beta_{att}$ under setting $\mathrm{VI}$ is given as
			\begin{align*}
	\varphi_{att, \mathrm{VI}} 
	={}& \frac{1}{qe_1}  \left ( \frac{A(Y - \mu_1(X))  }{ \P(A=1|X) }  -  \frac{(1-A)(Y - \mu_0(X)) }{ \P(A=0 |X) } \right )  \pi(X) e_1(X)   + \frac{G A}{q e_1} \left ( \mu_1(X) - \mu_0(X) - \beta_{att} \right ). 
	\end{align*}
The semiparametric efficiency bound is $\mathbb{\bar  V}_{att, \mathrm{VI}}^* =  \E[\varphi_{att, \mathrm{VI}}^2]$.		
	In particular, 
when $e_0(X) = 0$, $\varphi_{att, \mathrm{VI}}$ reduces to $\varphi_{att, \mathrm{V}}$; 
When $\pi(X) = 1$, $\varphi_{att, \mathrm{VI}}$ reduces to $\varphi_{att, \mathrm{I}}$, which is the EIF based only the source data. 
\end{theorem}

\begin{theorem}[efficiency comparison of $\beta_{att}$] \label{thm4-suppl} We have that 

(a) the semiparametric efficiency bounds of $\beta_{att}$ under settings $\mathrm{I}$-$\mathrm{VI}$ satisfy that 
    \begin{align*}  
     \mathbb{\bar V}_{att, \mathrm{I}}^* ={}& \mathbb{\bar V}_{att, \mathrm{II}}^*  =\mathbb{\bar V}_{att, \mathrm{III}}^* =\mathbb{\bar V}_{att, \mathrm{IV}}^*  >  \mathbb{\bar V}_{att, \mathrm{V}}^*, \\
     \mathbb{\bar V}_{att, \mathrm{I}}^* ={}& \mathbb{\bar V}_{att, \mathrm{II}}^*  =\mathbb{\bar V}_{att, \mathrm{III}}^* =\mathbb{\bar V}_{att, \mathrm{IV}}^*  > \mathbb{\bar V}_{att, \mathrm{VI}}^*. 
    \end{align*}

  (b) the efficiency gain of setting $\mathrm{V}$ compared to setting I is 
    \begin{align*}
    \mathbb{\bar V}_{att, \mathrm{I}}^* - \mathbb{\bar V}_{att, \mathrm{V}}^* =  \frac{1}{q^2 e_1^2}  \E \Biggl [ \pi(X) e_1(X)^2 \gamma(X) \frac{\text{Var}( Y(0) | X)}{1 - e_1(X)}  \Biggr  ], 
\end{align*}
where $\gamma(X)$ is defined in Theorem \ref{thm2}(b). 


(c) the efficiency gain of setting $\mathrm{VI}$ compared to setting I is
\begin{align*}
    \mathbb{\bar V}_\mathrm{I}^* - \mathbb{\bar V}_{\mathrm{VI}}^* = \frac{1}{q^2 e_1^2}  \E \Biggl [ \pi(X)^2  e_1(X)^2 \Big \{  \alpha_1(X)  \cdot \frac{\text{Var}( Y(1) | X)}{e_1(X) }   + \gamma_1(X) \cdot  \frac{ \text{Var}( Y(0) | X) }{(1 - e_1(X) )} \Big \} \Biggr  ],
\end{align*}
where $\alpha_1(X)$ and $\gamma_1(X)$ are defined in Theorem \ref{thm2}(c).

(d) the difference between $\mathbb{\bar V}_\mathrm{V}^*$ and $\mathbb{\bar V}_{\mathrm{VI}}^*$ is given as 
 \begin{align*}
    \mathbb{\bar V}_\mathrm{V}^* - \mathbb{\bar V}_{\mathrm{VI}}^*
   ={}&  \frac{1}{q^2 e_1^2}  \E \Biggl [ \pi(X)^2  e_1(X)^2   \Big \{  \alpha_2(X) \cdot \frac{\text{Var}( Y(1) | X)}{\tilde \P(A=1|X) }  - \gamma_2(X) \cdot  \frac{ \text{Var}( Y(0) | X) }{ \tilde \P(A=0|X) } \Big \} \Biggr  ]. 
\end{align*}
where  $\alpha_2(X)$ and $\gamma_2(X)$ are defined in Theorem \ref{thm2}(d).
	 		 
\end{theorem}

\section{Discussion}
 \label{sec:conclusion}
In this paper, we explore various scenarios with differing data structures and assumptions, followed by a comprehensive theoretical and empirical comparative analysis. We first conduct a comparative analysis by deriving the semiparametric efficiency bounds for the ATE in the target population under covariate shift, and then extend this analysis to new settings by further introducing posterior drift. Our conclusions emphasize the key factors influencing efficiency gains and losses, particularly the roles of the variance ratio of potential outcomes between datasets and the posterior drift functions in Assumption \ref{assump5}. We also consider sensitivity analysis by constructing the EIF based estimators for each sensitivity function to explore the estimation robustness.

Our proposed method may be extended to multiple source and target datasets \citep{Han-etal-2023, Han-etal2023-NIPS, Rosenman2023, Yang-Ding2020, Chen-Cai-2021}.  For these types of problems, the current estimation methods can be summarized into a two-step framework:  (1) Construct multiple estimators for the target estimand based on each individual source/target dataset or combinations of these datasets, using existing methods designed for one or two datasets. (2) Propose a strategy for combining the multiple estimators from Step (1). For example, \citet{Han-etal-2023, Han-etal2023-NIPS} used a regularization strategy, while \citet{Rosenman2023} employed a shrinkage strategy. An interesting direction for future work is to conduct comparative analysis involving multiple source datasets.

In addition, for cases where only partial covariates are observed in the target dataset \citep{Degtiar-Rose2023, Zeng-etal-2023}, especially when the covariates in the target dataset are a subset of those in the source dataset, there are two important issues worth further research. One is deriving and comparing the semiparametric efficiency bounds for the ATE in the target population under settings involving covariate shift and posterior drift \citep{Zeng-etal-2023}. The other is estimating the conditional average treatment effect curve for the target population and further exploring the estimation efficiency under different assumptions and data structures. The study of these issues is beyond the scope of this paper, and we leave them as topics for future research.

  
\bibliographystyle{apalike}
\bibliography{mybib}

\end{document}